%% file: NissenTree_v6_submP.tex
\documentclass[useAMS,usenatbib]{mn2e}

\usepackage{graphicx}
\usepackage{amsmath}
\usepackage{amsfonts}
\usepackage{amssymb}
\usepackage{multirow}
\usepackage{times}
\usepackage[T1]{fontenc}
\usepackage{aecompl}
\usepackage[normalem]{ulem}
\usepackage{longtable}
\usepackage{multicol}
\usepackage{blkarray}
\usepackage{txfonts}
\usepackage{color}
\usepackage[colorlinks=true, linkcolor=red, urlcolor=blue, citecolor=blue]{hyperref}

\newcommand{\tab}[1]{Table~\ref{#1}}
\newcommand{\fig}[1]{Fig.~\ref{#1}}
\newcommand{\sect}[1]{Sect.~\ref{#1}}
\newcommand{\sn}{solar neighbourhood}
\newcommand{\ymg}{$\mathrm{[Y/Mg]}$}

\def\kms{\,\mathrm{km\,s}^{-1}}
\def\kpc{\,{\rm kpc}}

\begin{document}

\title[Phylogeny in the solar neighbourhood]{Cosmic phylogeny: reconstructing the chemical history of the solar neighbourhood with an evolutionary tree}

\author[Jofr\'e et al.]{Paula Jofr\'e$^{1,2}$\thanks{E-mail: pjofre@ast.cam.ac.uk}, Payel Das$^3$, Jaume Bertranpetit$^{4,5}$ \& Robert Foley$^{5}$  \\
$^{1}$Institute of Astronomy, University of Cambridge, Madingley Road, CB3 0HA, UK\\
$^{2}$N\'ucleo de Astronom\'ia, Facultad de Ingenier\'ia, Universidad Diego Portales,  Av. Ej\'ercito 441, Santiago, Chile \\
$^{3}$Rudolf Peierls Centre for Theoretical Physics, University of Oxford, OX1 3NP, UK\\
$^{4}$Institut de Biologia Evolutiva, Universitat Pompeu Fabra, 08002, Barcelona, Spain\\
$^{5}$Leverhulme Centre for Human Evolutionary Studies, Department for Anthropology and Archaeology, University of Cambridge, CB2 1QH, UK\\
}
\date{Accepted .... Received ...; in original form ...}


\maketitle

\label{firstpage}

\begin{abstract}
Using 17 chemical elements as a proxy for stellar DNA, we present a full phylogenetic study of stars in the \sn. This entails applying a clustering technique that is widely used in molecular biology to construct an evolutionary tree from which three branches emerge. These are interpreted as stellar populations which separate in age and kinematics and can be thus attributed to the thin disk, the thick disk, and an intermediate population of probable distinct origin.  
We further find six lone stars of intermediate age that could not be assigned to any population with enough statistical significance. Combining the ages of the stars with their position on the tree, we are able to quantify the mean rate of chemical enrichment of each of the populations, and thus show in a purely empirical way that the star formation rate in the thick disk is much higher than in the thin disk. We are also able to estimate the relative contribution of dynamical processes such as radial migration and disk heating to the distribution of chemical elements in the \sn. Our method offers an alternative approach to chemical tagging methods with the advantage of visualising the behaviour of chemical elements in evolutionary trees. This offers a new way to search for `common ancestors' that can reveal the origin of {\sn} stars.
\end{abstract}

\begin{keywords}
phylogeny -  stars: solar-type - Galaxy: solar neighbourhood - Galaxy: evolution - methods: statistical 
- methods: data analysis
\end{keywords}

\section{Introduction}
\label{sec:Introduction}

In 1859, Charles Darwin published his revolutionary view of life, claiming that all organic beings that have ever lived have descended from one primordial form \citep{darwin2003origin}. One important outcome of Darwin's view of `descent with modification' was the recognition that there is a `tree of life' or {\it phylogeny} that connects all forms of life. The key assumption in applying a phylogenetic approach is that there is continuity from one generation to the next, with change occurring from ancestral to descendant forms. Therefore where two biological units share the same characteristics they do so because they have normally inherited it from a common ancestor. 

This assumption is also applicable to stars in galaxies, even if the mechanisms of descent are very different. 
We know that population I, II and III stars emerged from  a gas cloud whose primordial composition has evolved with time - in other words, population I stars were made from matter present in population II stars, and population II stars from matter in population III stars.  
Broadly speaking, the most massive stars explode in supernovae (SNe) donating metal-enriched gas to the interstellar medium (ISM), which eventually accumulates to form new molecular clouds and produce a new generation of stars. This cycle has been repeating ever since. Less massive stars {  ($M < 0.8 M_\odot$) } live longer than the age of the Universe and therefore serve as fossil records of the composition of the gas at the time they formed. Two stars with the same chemical compositions are therefore likely to have been born in the same molecular cloud. This process of `descent' mirrors that of biological descent, even though biological evolution is driven by adaptation and survival, while chemical evolution is driven by mechanisms that lead to the death and birth of stars. In other words,  here the shared environment is the ISM, which may be the key aspect studying evolution, rather than shared organism heredity key in biology. This raises the question of whether biologically-derived techniques for reconstructing phylogeny can be applied to galaxy evolution. 

Phylogenetic techniques have existed as long as evolutionary biology, but the strength of phylogenies based on DNA rather than phenotypes (aspects of morphology or biological structures, such as skulls), is that the mechanisms of change can be quantified, and so rates of change in DNA can be estimated independently \citep{lemey2009phylogenetic}. In astrophysics the chemical pattern obtained from spectral analysis of FGK type stars can be interpreted as {\it stellar DNA}, as it remains intact for the majority of their lives  \citep{2002ARA&A..40..487F}. The mechanisms for change in chemical abundances can also be identified. There is enrichment of the ISM, which is relatively well understood due to advances in nucleosynthesis and SNe yield calculations \citep{mcwilliam+04,mattuecci+12, 2011ApJ...729...16K}. Differences in chemical abundances of stars can also be the result of environmental processes bringing gas and stars from extragalactic systems and dynamical processes. Dynamical processes are a result of perturbations from nearby non-axisymmetric features such as the bar, spiral arms, molecular clouds, or merger activity. This can lead to radial migration, which is a change in the angular momentum that conserves the orbit's eccentricity or heating, which is a change in the eccentricity that conserves the angular momentum \citep{sellwood+02, 2010ApJ...722..112M}. Therefore abundance gradients produced from radial gas flows can result in metal-richer stars born in the inner Galaxy and metal-poorer stars born in the outer Galaxy being brought into the \sn. Fossils of the same chemical enrichment history born at different epochs should show a direct correlation between chemical difference and age, and therefore knowledge of stellar ages can help quantify the balance between the two possibilities. Kinematic information can then further help distinguish between origins of stars arising from different chemical enrichment histories.

{ Clustering algorithm leading to t}rees have already been { developed} in astrophysics with { a} method called {\it astrocladistics} \citep{2006A&A...455..845F, 2009MNRAS.398.1706F, 2015MNRAS.450.3431F}. In one application they { relate the morphology of} dwarf galaxies of the Local Group \citep{2006A&A...455..845F}, finding three groups { emerging from one common ancestor}. The second application was to study stellar populations in $\omega$-centauri  \citep{2015MNRAS.450.3431F} { finding} three populations with distinct chemical, spatial, and kinematical properties that they believe originate from gas clouds of different origins.  

{ While these works have shown the advantages of using clustering algorithms that are  also employed in biology in astrophysics, they have not explored in full the power of using trees. A tree can give us an extra dimension to the clustering algorithm: history.  As shown and discussed in the papers mentioned above, astrocladistics essentially attempts to represent the pattern of morphological similarity in e.g. dwarf galaxies. Phylogenetics, on the other hand, tries to represent the branching pattern of evolution \citep[see for instance][for further discussion]{ridley1986evolution}. A full phylogenetic analysis essentially consists in weighting shared characteristics according to the depth of the shared ancestry.  Thus, there is a subtle but very important difference between the work of astrocladistics and the goal of this paper. This is also the subtle difference between any other chemical tagging work and the goal of this paper} 

Here we  perform a full {\it phylogenetic} study of a sample of \sn\ stars, { in other words, here we attempt to study the evolution of the \sn\ by using a sample of stars and tree thinking}. We aim to obtain an insight into the chemical enrichment history of stars in different components in the Milky Way, using chemical abundance ratios as stellar DNA. { To this end we need to employ a different method to the } work of \cite{2015MNRAS.450.3431F} {  to construct the tree}. Here we use genetic distance methods (see \sect{method}) and not the maximum parsimony method. Maximum parsimony methods are designed to do cladistics, i.e, to determine the branching sequence and classify families of organisms. Distances methods, on the other hand, enable a full phylogenetic study to be conducted by analysing the branch lengths in the tree in detail. This allows us to understand how chemical enrichment history varies between the identified stellar populations. Our application therefore offers a complementary approach to astrocladistics as it allows us to go beyond a pure phenomenological classification and study evolutionary processes.

\section{Data}\label{data}
 
We demonstrate the phylogenetic approach on a sample of solar twins, for which accurate abundance ratios for several elements have been derived by \cite{Nissen15, Nissen16}. The sample is selected from FGK stars with precise effective temperatures, surface gravities and metallicities derived from spectra measured with the High Accuracy Radial velocity Planet Searcher (HARPS). This is a high-resolution ($R\sim 115000$) echelle spectrograph at the European Space Observatory (ESO) La Silla 3.6m telescope. The sample was selected to have parameters within $\pm100\,$K in $T_{\mathrm{eff}}$, $\pm0.15\,$dex in $\log g$, and $\pm0.10\,$dex in [Fe/H] of those known for the Sun. The sample is comprised of 22 stars: 21 solar twins and the Sun. Abundances are available for C, O, Na, Mg, Al, Si, S, Ca, Sc, Ti, Cr, Mn, Cu, Fe, Ni, Zn, Y and Ba, with typical measurement errors of the order of 0.01~dex. We consider the abundances in the $\mathrm{ [X/Fe]}$ notation
 and thus, the Sun is assumed to have $\mathrm{[X/Fe]} = 0$. 

There are several advantages to using such a sample. It decreases the systematic uncertainties associated with determining abundances for stars of different stellar spectral types. 
Furthermore, the stellar parameters and chemical abundance measurements for solar twins were done differentially with respect to the Sun. Differential analyses are well-known to significantly increase the accuracy of results with respect to direct absolute abundance measurements \citep[see e.g.][]{2006ApJ...641L.133M, 2014MNRAS.439.1028D, 2015A&A...582A..81J, 2016arXiv160604842S}. A related advantage, pointed out by \cite{Nissen15} and later extensively discussed by \cite{2016arXiv160604842S} is that the well-constrained stellar parameters allow  accurate ages (with errors $\lesssim 0.8\,$Gyr) to be determined from isochrones as they only span a restricted area in the HR diagram. This enables us to consider stellar ages in our interpretation (\sect{agesdynprop}). 

We also note that choosing a particular spectral type introduces a selection bias, which has to be taken into account when interpreting our evolutionary tree. For example, using solar metallicities will bias the sample towards thin-disk stars rather than thick-disk and halo stars. Therefore in this study we can not perform a quantitative analysis of the number of stars in different populations. This sample is still suited for our analysis since this small slice in metallicity still hosts a range of abundances and ages ($0.7$ to $9.8$ Gyr), as discussed by \cite{Nissen15}.

Finally, the abundances of the stars are supplemented with very accurate radial velocities derived from the cross-calibration procedure of the HARPS pipeline, which were downloaded from the ESO public archive\footnote{request number 235607}, as well as accurate astrometry  from Hipparcos. The astrometric data are shown along with the radial velocities in Table~\ref{astrometry}. A further comparison of these data with the new astrometric data from Gaia DR1 can be found in appendix~\ref{ap1}. We note these data are not used in the construction of the tree but will aid the interpretation of components that emerge in our evolutionary tree.

\section{Method}\label{method}
Here we discuss the steps involved in constructing the phylogenetic tree using the chemical abundance ratios, acquiring ages, and deriving dynamical properties of the stars.

\subsection{Tree construction}\label{constructtree}

Numerous tree-making methods have been proposed {  and discussed} in the literature \citep[see e.g.][]{sneath1973numerical, Felsenstein, felsenstein1988phylogenies, lemey2009phylogenetic, yang2012molecular}. In particular, genetic-distance methods are the appropriate ones for biological evolutionary studies, as the genetic difference between two organisms is directly related to the degree of evolution between them. The construction of a robust evolutionary tree in biology involves four main steps: (1) 
Define the set of categories  and the characters or traits that will be employed to determine shared ancestry and divergence; (2) construct a measure of the genetic distance between each pair of organisms to find the relation between them; (3) calibrate branch lengths to reflect the evolution of the system with time; and (4) assess the reliability of the tree topology. Each of these steps is explained below in more detail with relation to the application in our case. We recall that the array of abundance ratios is our stellar DNA sequence and therefore genetic distances will be hereafter called chemical distances.

\subsubsection{Categories}
These are commonly referred to in evolutionary studies as {\it Operational Taxonomical Units} (OTUs), or simply taxa. They can be either organisms of a given species, different species, or groups of species. OTUs are those at the end nodes (leaves) of the tree, i.e. they are the `observables'. Similarly, internal nodes in the tree are called {\it Hypothetical Taxonomic Units} (HTUs) to emphasise that they are the hypothetical progenitors of OTUs. Defining the set of taxa  is important as this will shape the evolutionary  analysis inferred from the tree.  In astrophysics, OTUs can be  individual stars, group of stars or dwarf galaxies as in the case of \cite{2006A&A...455..845F}. In this work we have 22 taxa, each representing one star of the sample described in \sect{data}.  

\subsubsection{Chemical distance matrix}
We define the chemical distance between the star $i$ and star $j$ as follows:
\begin{equation}\label{dist}
D_{i,j} =   \sum_{k=1}^{N} \big| \mathrm{[X}_k/\mathrm{Fe]}_i - \mathrm{[X}_k/\mathrm{Fe}]_j \big| \, ,
\end{equation}
\noindent  where the sum with respect to $k$ is over all abundance ratios $\mathrm{[X/Fe]}$. Each value $D_{i,j}$ comprises one element of the chemical distance matrix. The matrix is symmetric, with zeroes along the diagonal. This way of determining the chemical distance between stars is also used in the chemical tagging study of \cite{2013MNRAS.428.2321M}, but here we do not need to normalize by the number of chemical elements as this is always the same. We also do not explicitly consider weighted distances here, but they feature in determining the robustness of the tree in \sect{uncertainties}. { We comment here that this definition of chemical distance might be subject of degeneracies if the number of dimensions of the chemical space is large \citep[see ][for a discussion]{2015MNRAS.449.2604D}. We analyse this by considering a bootstrapping analysis (described below) which essentially serves to collapse any branch of a star that does not belong to a branch with enough statistical significance when different chemical elements are considered. We further comment that the chemical distance defined in Eq.~\ref{dist} can be replaced by other estimate of distance because is the clustering algorithm (next section) is what is the novel approach as it is the responsible to create the tree. A discussion of how this compares with other methods can be found in \sect{other_methods}.}

\subsubsection{Calculating branch lengths}
Several methods exist in the literature for calculating the branch lengths for the tree from the distance matrix. In biological evolution studies, the larger the genetic distance, the more evolution in general that separates two taxa and thus the larger the branch length. In the past, this evolutionary separation has been translated to time under the assumption of a `universal clock', i.e. the systems modify their characteristics at a constant rate. Recent studies have shown however that there is no universal clock, i.e. the rate of modification depends on location and taxa  \citep[see][for an extensive discussion]{lemey2009phylogenetic}. Therefore, the translation from genetic distance to time can only be directly done in very specific cases. In Galactic chemical evolution, chemical elements become more abundant from one generation to another, but they may do so at different rates in different Galactic components. Stars brought in from different radii due to radial migration or disk heating may also have followed different chemical enrichment histories. As with several cases in biology, we therefore need a method which allows the different branches to have different lengths, accounting for different enrichment rates.

In order to construct such a tree, we use the concept of minimum evolution, which is based on the assumption that the tree with the smallest sum of branch lengths is most likely to be the true one. \cite{Rzhetsky01091993} prove this tree has the highest expectation value, as long as the distance matrix used is statistically unbiased. As there are many possible trees that could be explored (see \sect{uncertainties}), we estimate the minimum-evolution tree using the widely-used neighbour-joining method \citep[NJ, ][]{Saitou01071987,Studier01111988}, which we explain below with regards to our application.\\ 

Consider five stars, $\mathrm{A}$, $\mathrm{B}$, $\mathrm{C}$, $\mathrm{D}$ and $\mathrm{E}$, which have the following chemical distance matrix with some arbitrary chemical abundance ratio units %
\[
\begin{blockarray}{cccccc}
		   &\mathrm{A} & \mathrm{B} & \mathrm{C} & \mathrm{D} & \mathrm{E} \\
\begin{block}{c(ccccc)}
\mathrm{A} & 0.0 & 5.3  & 4.6 & 7.1  & 6.1 \\
\mathrm{B} & 5.3 & 0.0  & 7.3 & 10.3 & 9.2 \\
\mathrm{C} & 4.6 & 7.3  & 0.0 & 7.5  & 6.2 \\
\mathrm{D} & 7.1 & 10.3 & 7.5 & 0.0  & 5.8 \\
\mathrm{E} & 6.1 & 9.2  & 6.2 & 5.8  & 0.0 \\
\end{block}
\end{blockarray} \, \,.
\] 

A traditional clustering method such as the nearest-neighbour algorithm would group stars $\mathrm{A}$ and $\mathrm{C}$ together as they have the shortest chemical distance between them. This would be correct if the evolution rate were the same everywhere. The NJ method allows different evolutionary rates by computing a rate-corrected chemical distance matrix, where the chemical distance between two stars is adjusted by subtracting the sum of the divergence rates of the two stars
\begin{equation}\label{dist2}
D^{\prime}_{i,j} =  D_{i,j} - (r_i + r_j).
\end{equation}
where $D_{i,j}$ are the elements of the chemical distance matrix defined in ~Equation \ref{dist}, $n^*$ is the number of stars, and the divergence rates are calculated as
\begin{equation}
r_j = \frac{\sum_{i\ne j} D_{i,j}}{(n^*-2)}.  
\end{equation}
The divergence rate is essentially the `mean' chemical distance  between each of the two stars and the other stars. We note that the denominator is $n^*-2$ rather than $n^*-1$ since we are considering $(n^*-1)$ stars for the star $j$. In our example, $r_2$ would be the divergence rate for star $\mathrm{B}$, with a value $(5.3+0.0+7.3+10.3+9.2)/3=10.8$. 

A new node, {\it ij}, is defined for the two stars for which $D^r_{i,j}$ is minimal, i.e. the chemical distance between them is small compared to how much those stars vary with every other star. The branch lengths from the new node {\it ij} to its `children' $i$ and $j$ are calculated as
\begin{equation}
D_{ij,i} = \frac{D_{i,j} + r_i-r_j}{2}.
\end{equation}
and similarly for $D_{ij,j}$. The total branch length between $i$ and $j$ is still just their chemical distance $D_{i,j}$. Now a new distance matrix is constructed with the new node {\it ij} in place of $i$ and $j$. This new distance matrix contains $(n^*-1)\times(n^*-1)$ elements. The chemical distances to the new node are calculated as
\begin{equation}
D_{ij,k} = \frac{D_{i,k} + D_{j,k} - D_{i,j}}{2} .
\end{equation}
If the element of the new matrix for which the rate-corrected chemical distance is smallest is between the new {\it ij} node and another star, say C, a new node is created that branches into the {\it ij} node and C. Otherwise the new node is created at the end of a new branch. The whole process is repeated until the tree topology is fully resolved. We only have dichotomies. The advantage of this method is that it is very fast, which is crucial when dealing with a large sample of taxa and when assessing the statistical significance of the tree (\sect{uncertainties}). Furthermore, this is a widely-used method in biology, and therefore many implementations exist as packages. We use the NJ method implemented in the software {\tt MEGA} v7.0\footnote{http://www.megasoftware.net/home} \citep{kumar2016mega7}. This software, unlike many other codes, is able to construct trees from pre-defined distance matrices, rather than the original DNA sequences.  Furthermore, {\tt MEGA} can be easily called from Python which is important for studying the robustness of the tree. 

\subsubsection{Assessing robustness}\label{uncertainties}
Inferring an evolutionary tree is an estimation procedure, in which a `best estimate' of the evolutionary history is made on the basis of incomplete information. In the context of molecular phylogenetics, a major challenge is that many different trees can be produced from a set of observables. With 20 taxa for example, there are close to $10^{22}$ trees of the type we have constructed \citep{dan2000fundamentals}. With current spectroscopic surveys of Milky Way stars, we can have millions of stars with detailed chemical abundances, producing an enormous number of tree possibilities. Here we only use 22 solar twins, but this still implies a huge number of possible trees. Of course, many of these have very low probability and therefore it is very important to develop a statistical approach to infer the most probable trees among this huge sample. Even if we use the NJ method to help estimate the most likely tree, measurement errors in the abundance ratios and systematic errors in our choice of elements to represent the chemical sequence will distort the likelihood distribution. It is therefore crucial before interpreting the tree and reconstructing the evolutionary history of our sample of stars to ensure that our final tree is robust by means of statistical techniques \citep[see e.g.][for a discussion]{felsenstein1988phylogenies}.  

Here, our robust tree was obtained by combining a Monte Carlo method that explores measurement error with a bootstrapping procedure that explores alternative definitions of the chemical sequence:
\begin{enumerate}
\item {\it Monte Carlo simulations: } 
This is a widely-used method for propagating uncertainties in measured quantities. The measurement errors in the abundance ratios are of the order of 0.01~dex. Using those listed in Table~4 of \cite{Nissen15} and Table~2 of \cite{Nissen16}, we generate new data matrices by drawing from Gaussian distributions with means equal to the abundance ratios in the old data matrix and dispersions equal to the measured errors. In doing this we assume the errors in [Fe/H] are very small and therefore essentially independent.\\
\item {\it Bootstrapping/jackknifing: }
This is commonly employed in evolutionary studies, where random characters in the DNA sequence are removed from the data matrix. As our chemical sequences are significantly shorter than DNA sequences --17 abundance ratios in comparison to millions of genes -- we instead create alternative chemical sequences in which the length of the chemical sequence is kept constant at 17 using the new data matrix generated by the Monte Carlo method. These are created by generating a random sequence of the original 17 abundance ratios, using sampling with replacement. Thus we obtain chemical sequences that may not contain all the elements, and the elements that are included will have a random weight. 
\end{enumerate}
1000 resamples of the original data matrix are produced using the combined Monte Carlo and bootstrapping method, each consisting of 22 stars and a chemical sequence of length 17. New chemical distance matrices and trees are calculated for each resample. A consensus tree is then built from the 1000 trees using the software {\tt Mesquite v3.10}\footnote{http://mesquiteproject.org/}. A majority-rule consensus method is used for which branches only appear that are statistically significant, i.e. that appear in a non-conflictive way in at least 50\% of the trees. The majority-rule consensus method is particularly important for assessing the significance of very short branches. If these branches are not statistically significant, they are removed and  polytomies (more than two branches extending from one node) are obtained rather than dichotomies (two branches extending from one node). This is important because a fundamental concept in evolutionary biology studies is that natural processes divide a population into 2, enabling further independent evolution in dichotomies.  Polytomies, on the other hand, are interpreted as a burst of diversification: many species diverging at the same time, with a single origin. In biology this is rather unlikely, and therefore polytomies are treated with caution and interpreted as unresolved data or incomplete data \citep{felsenstein1988phylogenies}. In Galactic chemical evolution however, polytomies are expected consequences of star formation bursts and therefore constructing the majority-rule consensus tree is a fundamental step of applying phylogenetics in our case.

{ We comment that Monte Carlo simulations are standard  methods to assess the robustness of trees in evolutionary biology.  These are combined with bootstrapping by adding and removing genes (chemical elements in our case).  There might be other approaches to do this, but they do not seem in biology to be competitive enough to replace the standard MC and bootstapping methods. 
Some more sophisticated methods have been proposed in the literature, like the double bootstrap \citep[e.g.][for recent discussion]{ren2013assessing}. These methods however do not significantly improve results with respect to standard methods but lower down the computational time to obtain final trees, which is relevant when large datasets are being analysed.  A discussion of this can be also found in \cite{ropiquet2009supertri}. 

It is true that in astrophysics this could be different, but here we are trying to make the first step, which is to use what are the standard techniques in biology.  Since we do not find anything conflictive in our results (below), we believe that applying this is sufficient for this first step. Certainly when numerical simulations and larger datasets are used,  more sophisticated techniques will have to be explored}.

\subsection{Ages and dynamical properties}\label{agesdynprop}
Knowledge of stellar ages will help explore how chemical distances correlate with time in identified stellar populations, and in the cases they do, quantify how the rates differ between them. The ages used in this work (\tab{actions_ages}) are taken from \cite{Nissen15, Nissen16}, which are derived considering the stellar parameters and isochrones. Combining ages with dynamical properties will help interpret the different branches that appear in the consensus tree. We calculate dynamical properties using the astrometric and kinematic information in \tab{astrometry}. To convert from heliocentric coordinates to Galactocentric coordinates, we assume that the Sun is located at $(R_0,z_0) = (8.3,0.014)\kpc$ \citep{schonrich+12}, that the local standard of rest (LSR) has an azimuthal velocity of $238\kms$, and that the velocity of the Sun relative to the LSR is $(v_R,v_{\phi},v_z) = (-14.0, 12.24,7.25)\kms$ \citep{schonrich+12}. 
\begin{figure*}
\hspace{-1cm}
\includegraphics[scale=0.5, angle=90]{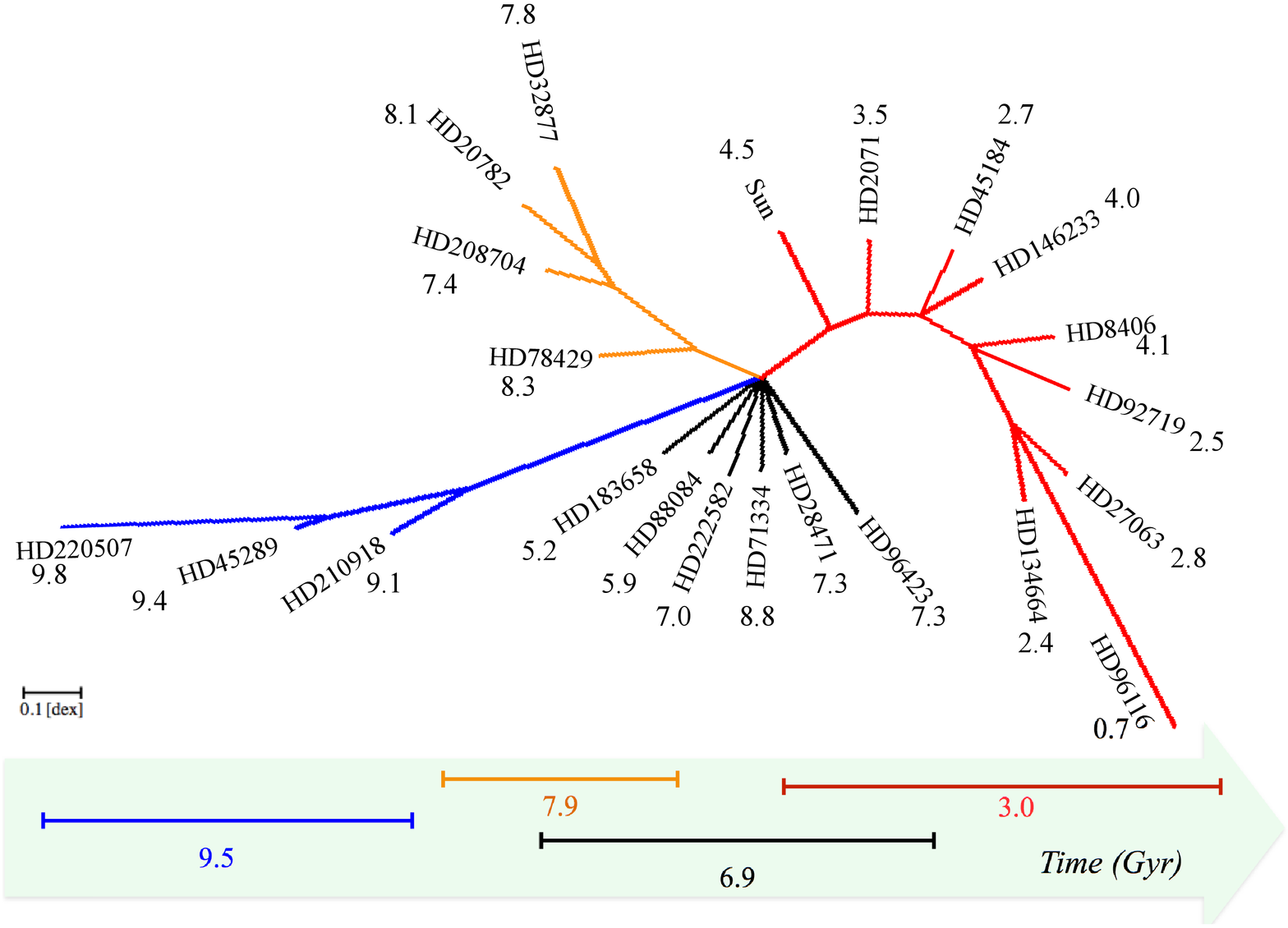}
\caption{Unrooted phylogenetic tree. Three main branches are obtained and coloured with blue, red and orange. Stars that do not belong to any branch with statistical significance are coloured with black. Branch lengths are in $\mathrm{dex}$ units, with the scale indicated at the bottom of the diagram.}
\label{radiation_tree}
\end{figure*}

We calculate the velocities ($U$,$V$,$W$) of the stars in heliocentric Cartesian coordinates, which tell us the velocities of the stars with respect to the Sun. We also calculate the actions $J_r$, $J_{\phi}$ and $J_z$ of the stars, which quantify the extent of the star's orbit in the radial, azimuthal and vertical directions. In an axisymmetric system, $J_{\phi}$ is simply the $z$ component of the angular momentum $L_{z}$. The actions will help identify which Galactic component the stars belong to as we would expect thin disk stars to be on near circular orbits in the equatorial plane and therefore have a small $J_r$ and $J_z$. The actions can be calculated from Cartesian coordinates using the St\"{a}ckel Fudge, given some gravitational potential assumed for the Galaxy. We use the composite potential proposed by \cite{dehnen+98}, generated by thin and thick stellar disks, a gas disk, and two spheroids representing the bulge and the dark halo. We use the potential parameters of \cite{piffl+14} and the implementation of the St\"{a}ckel Fudge by Vasiliev et al. (\textit{in prep.}), which is part of the \underline{A}ction-based G\underline{A}laxy \underline{M}odelling \underline{A}rchitecture (\texttt{AGAMA}\footnote{\texttt{AGAMA} can be downloaded from \\ \url{https://github.com/GalacticDynamics-Oxford/AGAMA}}). In calculating the velocities and actions of the stars, we propagate the errors in the astrometric and kinematic data using 1000 Monte Carlo samples to estimate the mean and dispersion. The dynamical properties can be found in \tab{actions_ages}.

\section{Results}\label{results}

\subsection{The consensus tree}\label{consensustree}
Figure~\ref{radiation_tree} presents our consensus tree. A tree constructed with the NJ method is `unrooted', and therefore even if clusters of stars appear to branch off initially from the same point, it does not signifiy this is the beginning of the evolution of system. The radiation format (\fig{radiation_tree}) for a phylogenetic tree is particularly suitable for visualising such trees as it emphasises that the groups and unclassified stars are separate from each other. The length of the branches are in chemical $\mathrm{dex}$ units with the scale indicated in the legend. They represent the chemical distance between each star and the node from which it emerged. The chemical distance between any two stars is determined by adding up the lengths of the branches between them.

The NJ method alone would only produce dichotomies, i.e. nodes split into two branches. However, the application of the majority-rule consensus method results in several polytomies, showing that some of the branches found in each of the 1000 trees in the resampling were not significant. Such polytomies could be expected in chemical evolution when star formation bursts can result in several stars driving different chemical enrichment paths,  not just two.

\begin{figure*}
\includegraphics[scale=0.38]{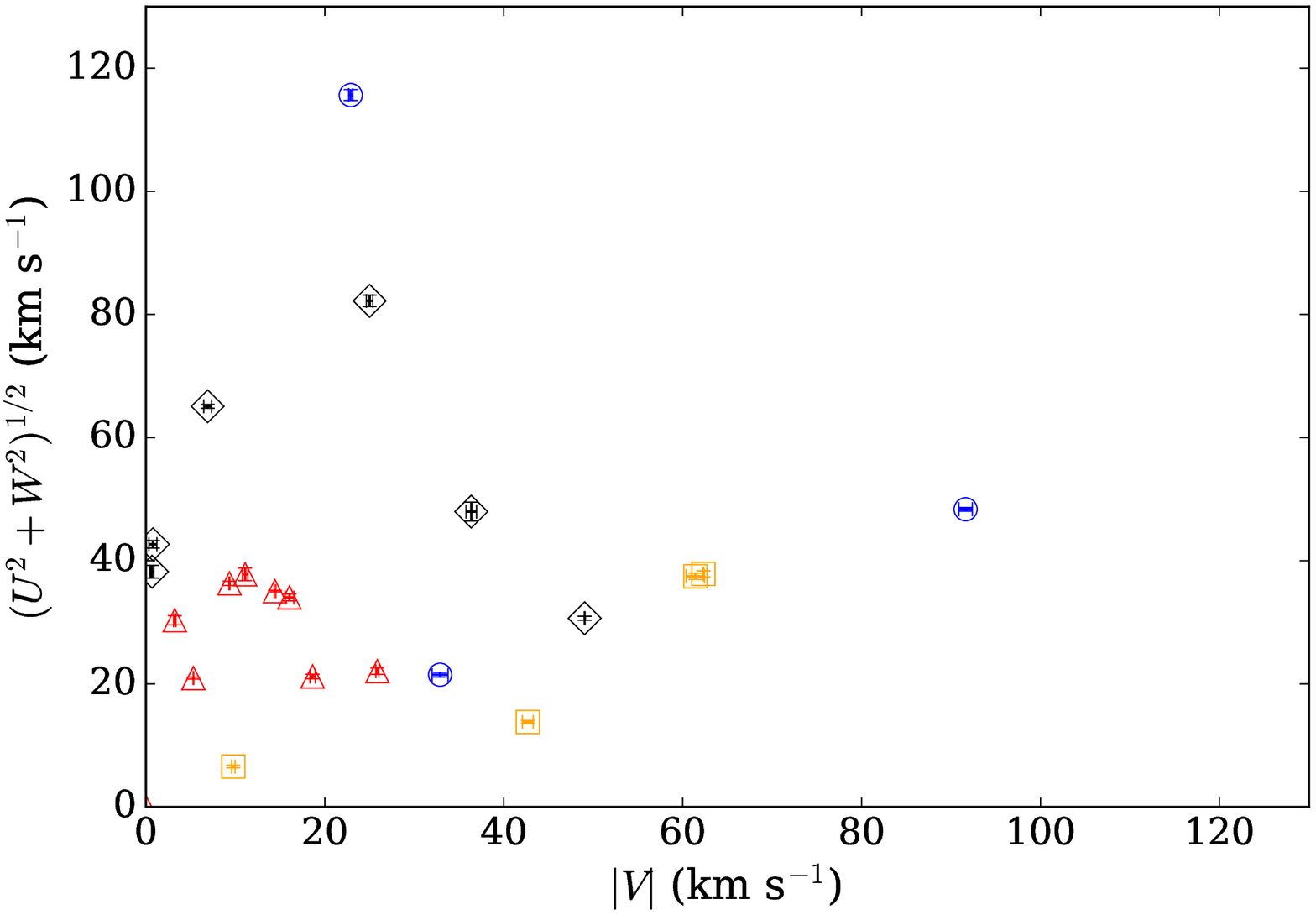}
\includegraphics[scale=0.38]{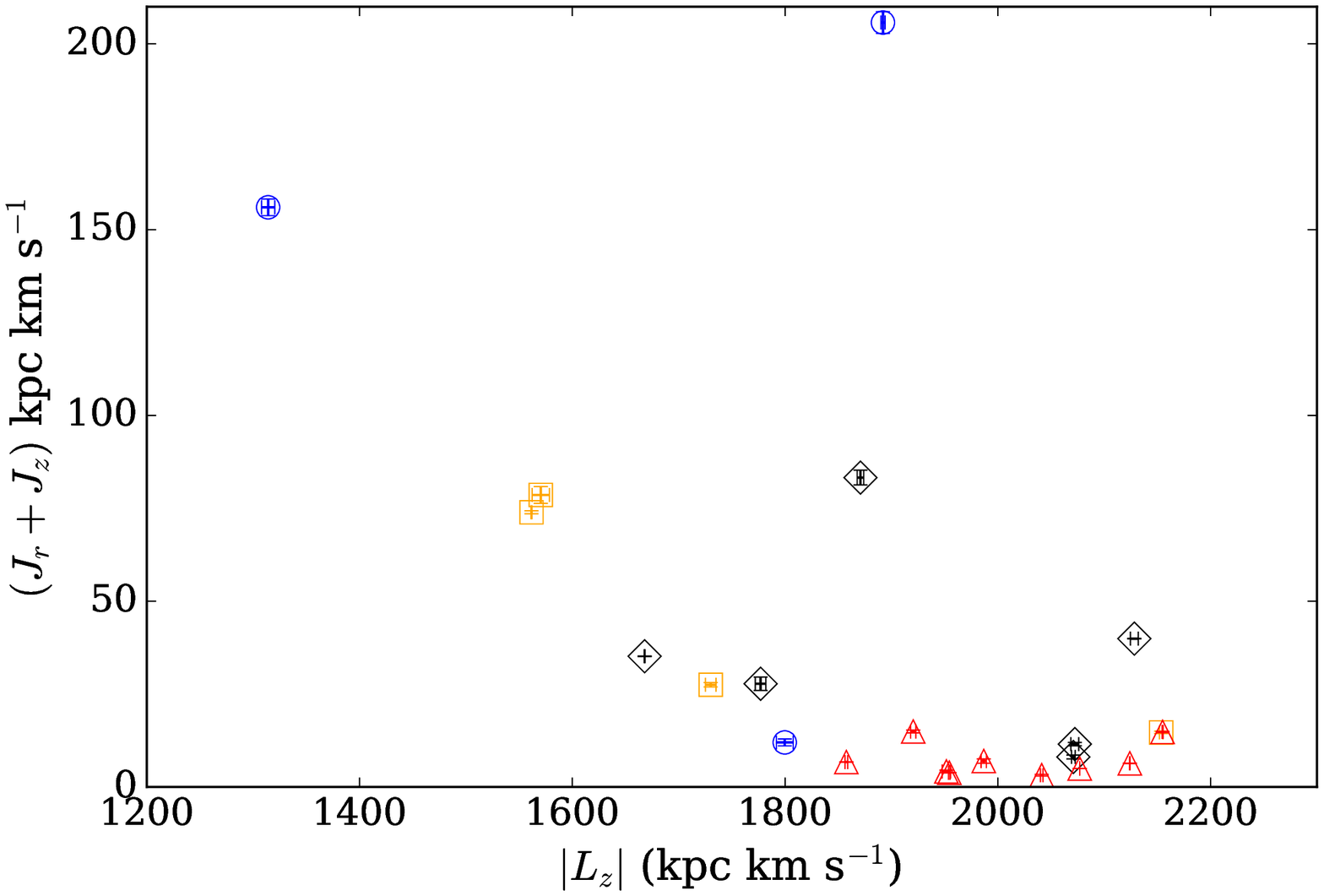}
\caption{Toomre diagram (left) and the sum of the radial and vertical actions shown against the $z$ component of angular momentum (right). Stellar populations are coloured according to the classification of the tree of \fig{radiation_tree}.}
\label{dynamical}
\end{figure*}

Even with the polytomies, three main stellar populations, i.e. groups of stars sharing a common ancestor, are identified. The first one (red), includes the following stars: the Sun, HD~2071, HD~45184, HD~146233, HD~8406, HD~92719, HD~27063, HD~96116 and HD~134664. A second stellar population (blue) includes the stars:  HD~210918, HD~45289 and HD~220507. A third stellar population (orange) appears to be equally independent from the other two populations, and includes the stars  HD~78429, HD~208704, HD~20782 and HD~38277. Finally, six stars (black) can not be assigned to any population with enough statistical confidence. These stars are HD~28471, HD~96423, HD~71334, HD~222582, HD~88084 and HD~183658.  { We comment here  that perhaps a different definition of chemical distance than the one employed in Eq.~\ref{dist} might help to allocate some of these 6 stars into one of the populations with better confidence but this is beyond the purpose of this work. Here we want to show that it is possible to apply phylogenetic analyses and tree thinking in the field of Galactic archaeology}. In this work we call these stars simply `undetermined'.

The branch lengths of the red stellar population are short, of the order of $0.1~\mathrm{dex}$, except for HD~96116, which has a branch length of almost $0.4~\mathrm{dex}$. The branch lengths of the orange stellar population are also of the order of $0.1~\mathrm{dex}$, reflecting that these stars are very chemically similar to each other.  The branch lengths between the original polytomy and the first node in the blue stellar population is significantly larger than the first nodes in the red and the orange populations, but the branch lengths between nodes is of the same order of magnitude, with the exception of HD~220507, which has a branch length of $0.3~\mathrm{dex}$.

For guidance, the ages of the stars are indicated next to their names in \fig{radiation_tree}. The star furthest along the red right branch, HD~96116, is the youngest star of the sample ($0.7~\mathrm{Gyr}$). It is more separate from the other stars in the red stellar population due to a larger chemical distance, and this is reflected in the ages of the other stars, which are clustered between $2.4$ and $4.5~\mathrm{Gyr}$. The orange stellar population has stars that are chemically very similar to each other. Their ages are also in a restricted range of $0.9~\mathrm{Gyr}$ between $7.4$ and $8.3~\mathrm{Gyr}$.

The star furthest along the blue branch, HD~220507, is also the oldest star of the sample ($9.8~\mathrm{Gyr}$). The other two stars in this branch have ages of $9.1$ and $9.4~\mathrm{Gyr}$. The chemical distance between the youngest and the oldest star is the largest in the tree at  $1.6~\mathrm{dex}$, and is obtained by adding the length of all the branches that separate the stars. The undetermined stars have ages between $5.2$ and $8.8~\mathrm{Gyr}$.

The mean age of all stars in a given population are indicated at the bottom of \fig{radiation_tree}, following the colour coding of the branches. We can see that the stellar populations have different ages, with time increasing from left to right. The ages have told us that there is a very old stellar population (blue branch),  a slightly younger but still old population (orange branch); stars of intermediate age  (black branches), and a young stellar population (red branch). We remark here that age was not used to build the tree. 

Thus,  the tree of \fig{radiation_tree} adds a rough direction to the evolutionary processes, i.e. the older stars are found towards the left, while the younger stars are found towards the right. However, it is clear that in the region of overlap between the red, black, and blue branches, the stars belonging to separate stellar populations may have been produced from gas that was chemically enriched at different rates and exposed to dynamical processes such as heating and radial migration. Therefore their location in this region is no longer indicative of their age. We will come back to the importance of dynamical processes in \sect{discussion}.

\subsection{Dynamics of the identified stellar populations}
In order to study the nature of the identified stellar populations more clearly, we look at dynamical properties of the stars. The left panel of \fig{dynamical} shows the classical Toomre diagram. The $y$ axis indicates the contribution to the star's velocity in directions perpendicular to that of Galactic rotation and the $x$ axis indicates the contribution to the star's velocity along Galactic rotation, compared to that of the Sun. Thin-disk stars should behave like the Sun, while thick disk and halo stars have velocities in random directions. Therefore the stars in the red population that cluster around zero in both directions, and include the Sun, behave like thin-disk stars. The other populations and undetermined stars cover a larger range in this plot, but in general tend to have higher values in both directions and are therefore more likely to be thick-disk stars. They are unlikely to be halo stars because they are comparatively metal-rich.

The right panel in \fig{dynamical} shows the `actions' equivalent of the Toomre diagram with the sum of the radial and vertical actions shown against the $z$ component of angular momentum, $L_z$. Thin-disk orbits are in the equatorial plane and should cluster in $L_z$, as we are restricted to the as \sn. The red stars have a low total radial and vertical action and cluster in $L_z$, and are therefore probably thin-disk stars. The blue stars show a range of total radial and vertical actions and a larger spread in angular momentum and therefore could be thick-disk stars. The other stars have a range of $L_z$ and low to intermediate total radial and vertical actions.

\begin{figure}
\includegraphics[scale=0.38]{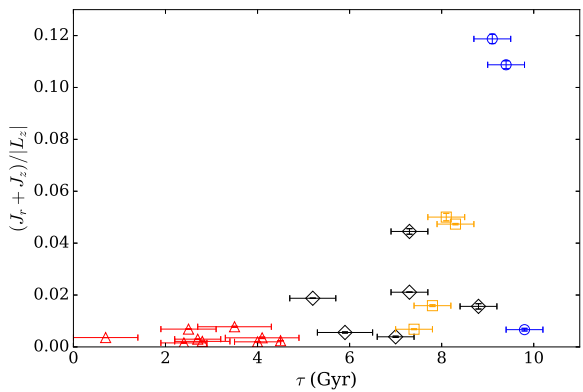}
\caption{Eccentricity against age of the stars for the stellar populations and undetermined stars identified in \fig{radiation_tree}.}
\label{ecc_age}
\end{figure}
The stellar populations in terms of ages and dynamical properties together are illustrated in \fig{ecc_age}. The $y$ axis combines the axes of the left panel of \fig{dynamical} into a dimensionless ratio of the sum of actions in the radial and vertical directions, divided by $L_z$. This dimensionless ratio is a measure of the eccentricity of the orbit \citep{sanders+16} as it normalizes the radial and vertical excursions of the star by the mean radius of the orbit. Thin-disk orbits are circular and therefore should have a small eccentricity while thick-disk orbits should have a range of eccentricities. This is plotted against the age. The red stars are most likely to be thin-disk stars as they have low eccentricities and have a spread in younger ages. The blue stars are most likely to be thick-disk stars as they are all old and have a range of eccentricities. The similarity between the orange stars is reflected clearly in their small range of ages. They have a range in eccentricities that is intermediate between the thin and thick disk.

The undetermined stars randomly span the ages between the thin and thick disks with a range of eccentricities intermediate between the thin-disk and thick-disk stars. Some of these stars could belong to the older, flared part of the thin disk and the oldest may belong to the thick disk. The small number of data points means however that if this were the case, the connection between these stars and those in the thin and thick disk could not be recovered. 

It should also be noted that the trend followed by the stars is continuous, restating the age-velocity dispersion relation \citep{wielen77}, i.e. older stars have a higher velocity dispersion and therefore lie on more eccentric orbits. There is no evidence from this plot that the data require separate components, and therefore a more careful inspection of the abundance ratios is required (see \sect{discussion}).

\subsection{Evolution of stellar populations}\label{evolution} 

\begin{figure*}
\hspace{-3.5cm}
\includegraphics[scale=0.6, angle = 90]{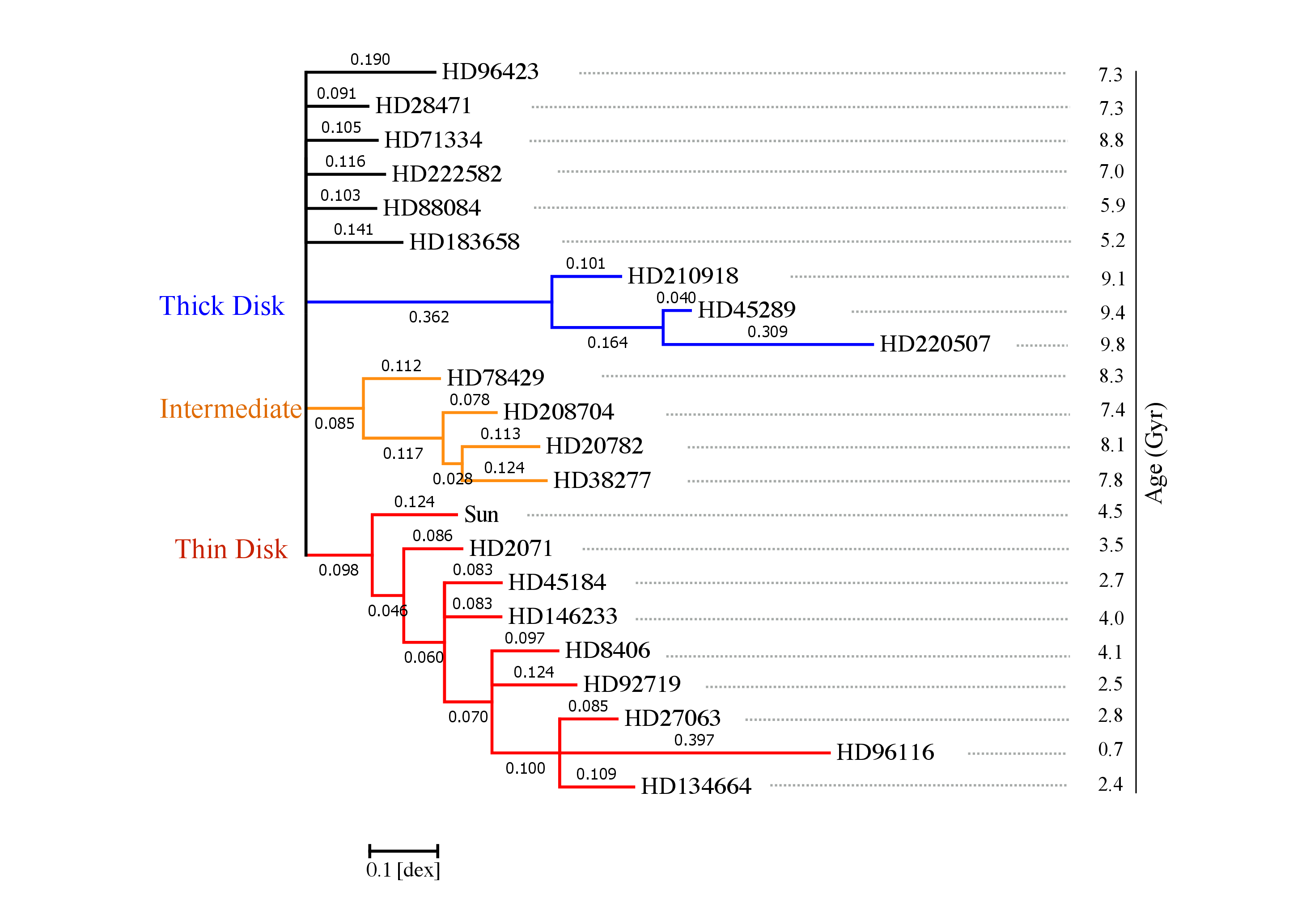}
\caption{Phylogenetic tree of 22 solar twins in the \sn, created using 17 elemental abundances. Stellar populations are assigned considering the age and the dynamical properties of the stars and are indicated at the right. Branch lengths have units in dex, with the scale indicated at the left bottom.}
\label{our_tree}
\end{figure*}

To go further than simply identifying stellar populations, the phylogenetic nature of the tree allows us to investigate the evolution of chemical abundances within each stellar population. For example, any time a branch splits (i.e. there is a dichotomy or polytomy), we can interpret this as an instance at which the evolution of the gas diverged. The stars arising from the split have formed from the same or similar gas clouds. The further away a star in the polytomy is from the node the more  evolved a gas cloud from which it originated. 

\begin{figure}
 \vspace{-0.5cm}
 {\hspace{-0.5cm}}
\includegraphics[scale=0.55]{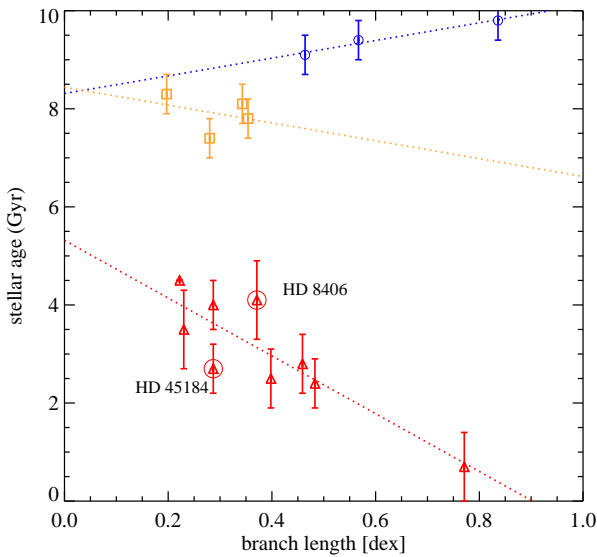}
\caption{Stellar age against branch length measured from the vertical edge from which the thin-disk branch (red), thick-disk branch (blue), and intermediate population branch (orange) emerge. }
\label{migration}
\end{figure}

We study the chemical evolution within these different Galactic components by exploiting the phylogenetic nature of the tree. To do so, we redraw the tree of \fig{radiation_tree} using a classical format in \fig{our_tree}, as this allows us to inspect the individual branch lengths in more detail. The Galactic components assigned to each branch are indicated on the left and stellar ages on the right for guidance. Since we focus here only on stars that are assigned to a branch with enough statistical relevance, we consider the blue, red, and orange branches independently, and calculate the total branch length starting from the vertical line. This is done by adding up the values indicated in each branch from the vertical line to each star. In \fig{migration} we show the branch-length and age (BLA) relation, finding that branch length decreases with stellar age for the thick-disk (blue stars) but increases with stellar age for the thin-disk and intermediate stars (red and orange stars). This is simply a consequence of the location of the zero point in the tree. We would expect environmental processes and dynamical processes that have brought stars from a very different part of the Galaxy to branch out as separate stellar populations. Therefore within each branch we would expect a monotonic trend of stars increasing their elemental abundances with time. However, stars brought in from radii near to the {\sn} may be allocated to the same branch, adding scatter to this trend.  We can therefore make an estimate of the mean chemical evolution rate and contributions from dynamical processes by carrying out linear fits (dashed lines in \fig{migration}). Below we discuss the chemical evolution history for the different groups of stars.\\

\noindent {\it Thin disk:} The BLA relation for the thin-disk stars is shown in red in \fig{migration}. The stars are not consistent with being a single age, and therefore the stars are likely to have formed in successive multiple bursts of star formation rather than a single burst. As suggested by the tree, the stellar age generally correlates approximately linearly with total branch length. Therefore we can make an estimate of the mean chemical evolution rate, $<\dot{\Sigma}_{XFe}>$, by adding up all the branch lengths between the node from which the Sun emerged and HD~96116 to obtain a total chemical distance of $0.673~\mathrm{dex}$. This distance emerged within a period of $4~\mathrm{Gyr}$, and therefore an indication of the mean chemical evolution rate can be obtained by dividing the total chemical distance by the difference in ages to get $<\dot{\Sigma}_{XFe}> = 0.168\,\mathrm{dex} \,\mathrm{Gyr}^{-1}$.

We also notice in \fig{migration}, that there is a significant degree of scatter about the straight-line fit of the BLA relation (i.e. assuming the mean chemical evolution rate is constant with time), which mean coeval stars show a different level of chemical evolution. Stars deviating more than $1\sigma$ from the fit are enclosed  with a circle, and their name indicated. 
These stars may be good candidates to have arrived to the \sn\ via radial migration because their eccentricities do not stand out from the rest of the thin-disk stars.  \\



\noindent {\it Thick disk:} The stellar ages against total branch length for the thick-disk stars are plotted in blue in \fig{migration}. The stars in this case are consistent with being a single age, and therefore could have formed from a single burst, but the diversification in chemical sequences suggests they were not all formed at exactly the same time. Therefore the burst was not instantaneous. As with the thin-disk stars, the BLA relation is linear, although we must emphasise that in this case the data are also consistent with a zero-gradient curve. Making a simple estimate of the chemical evolution rate as with the thin disk, we obtain a total chemical distance of $0.473~\mathrm{dex}$ and a total age span of $0.7~\mathrm{Gyr}$. This gives a rate of $<\dot{\Sigma}_{XFe}> 0.657 \,\mathrm{dex} \,\mathrm{Gyr}^{-1}$, i.e. a higher rate compared to the thin disk.\\

%
\noindent {\it Intermediate:} The orange stars of \fig{migration} lack a clear evolution direction, i.e. the BLA relation is not evident. The stellar ages can also be seen to be relatively consistent with each other within the errors. Therefore, like the thick disk, these stars could have formed in a single burst and again the small level of diversification in chemical sequences suggests  that the burst was not instantaneous. Despite the lack of a clear direction of chemical evolution, we can still place a lower limit on the chemical evolution rate by considering that a total branch length of $0.269~\mathrm{dex}$ was experienced within $1~\mathrm{Gyr}$, giving a lower limit on the chemical evolution rate of $<\dot{\Sigma}_{XFe}> 0.269\,\mathrm{dex} \,\mathrm{Gyr}^{-1}$, which is in between both populations.\\

\begin{figure*}
\includegraphics[scale=0.38]{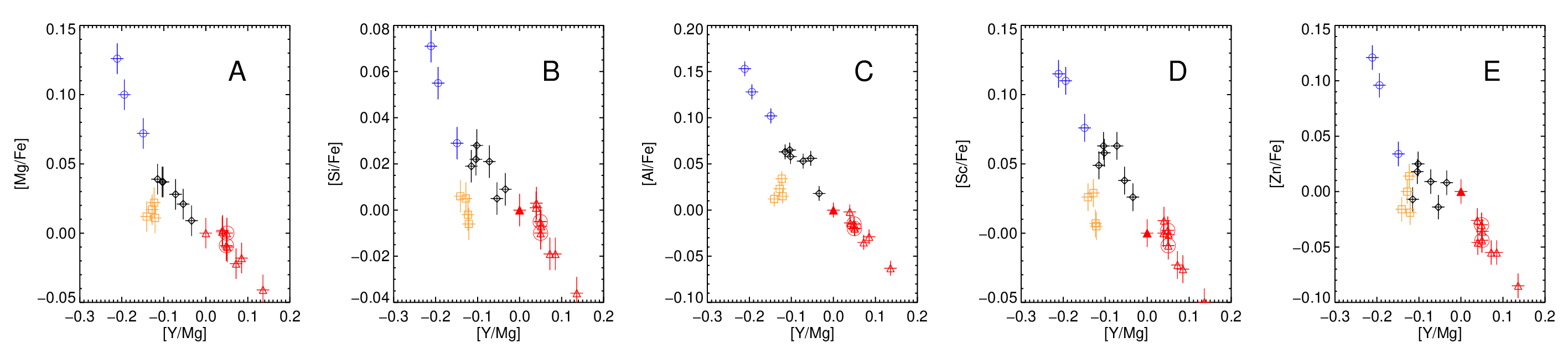}
\includegraphics[scale=0.38]{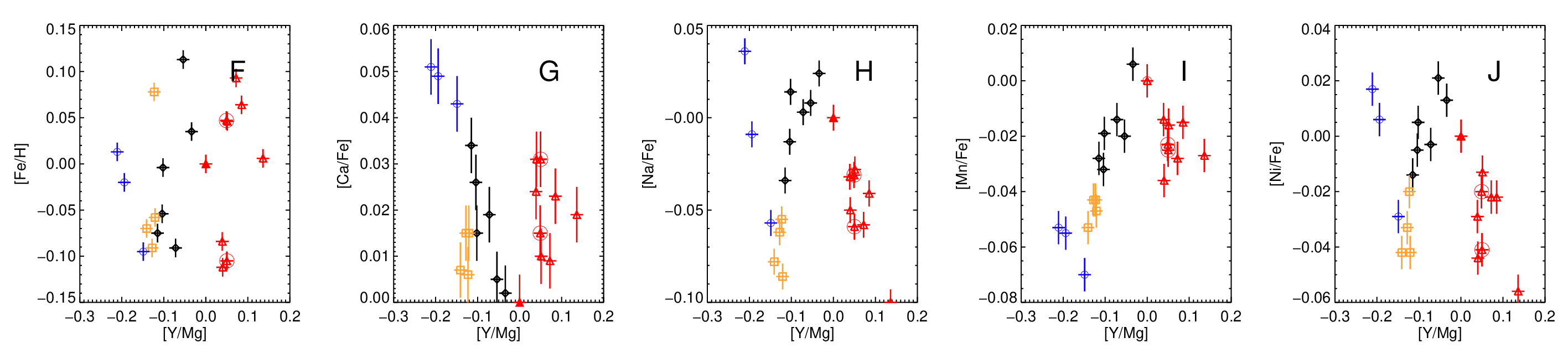}
\caption{Abundance ratios as a function of \ymg\ for the stellar populations and undetermined stars identified in \fig{radiation_tree}. The abundance ratios in the upper raw of panels (Mg, Si, Al, Sc and Zn) show a strong correlation of [X/Fe] with \ymg\ for the bulk of the populations. The abundance ratios in the lower raw of panels (Fe, Ca, Na, Mn and Ni) do not show a clean correlation versus \ymg. Stars enclosed correspond to those enclosed in \fig{migration}.}
\label{pops}
\end{figure*}

%
\noindent {\it Undetermined:} The black stars do not form a branch themselves in \fig{our_tree}, and therefore we can not include them in \fig{migration}. Nonetheless, they are interesting given that they emerge from the central polytomy of the tree. As the tree is not rooted, the undetermined stars are not connected to each other and therefore either formed in separate bursts from each other and the other stellar populations or could not be connected to the other stars. They all have similar chemical distances of 0.1 dex compared to the stars in the other stellar populations, and cover a more extended age range of 3 Gyr with respect to the intermediate or the thick-disk populations. As they form a polytomy, this could indicate a more extreme event that occurred around 7 Gyr ago, triggering a separation of the chemical evolution of the disk into several paths. Alternatively, some of these stars could be accreted stars, or just formed from the gas clouds related to those from which the other stellar populations emerged. A larger dataset of solar twins, analysed in the same way as the stars used here, is needed to understand them better. 


\section{Discussion}\label{discussion}
\subsection{Are the stellar populations identified in the tree really distinct?}
%
\cite{Nissen16} studied the abundance ratios as a function of age for all the elements studied in this work, showing that stars with ages below $6~\mathrm{Gyr}$ (which in our case would include the thin-disk stars) have a distinct behaviour from older stars. More specifically, his Fig.~3 shows how the $\alpha-$enhanced stars (our thick-disk stars) have a notably different trend of [X/Fe] as a function of age with respect to the young stars for Na, Sc, Cr, Mn, Cu, Ni and Zn. While an extensive discussion of these trends can be found in that paper, here we use them to argue that although thin and thick disk stars might show a continuous behaviour in their dynamical properties as a function of age, some of the abundance ratios suggest that these two populations might be distinct. 

In the upper panel of \fig{pops} we show  the abundance ratios of  Mg, Si, Al, Sc, and Zn over Fe as a function of \ymg, a proxy for age \citep{Nissen16}\footnote{See recent discussion of \cite{2016arXiv161003852F} regarding how \ymg\ is good proxy for age for main-sequence stars of solar metallicities but not so good for lower metallicitites.}, with the stars colour-coded by their assigned population.  These ratios evolve smoothly with \ymg\ in the case of the thick-disk, the undetermined and the thin-disk stars, with no obvious gap in chemical space. The abundance ratios of Fe, Ca, Na, Mn and Ni are shown as a function of \ymg\ in the lower panel of \fig{pops}. Several abundances show an offset in their dependence with \ymg\ that is difficult to reconcile between the various populations. The offset in Zn of 0.15~dex between the thin and thick disks has already been discussed by e.g. \cite[and references therein]{2015A&A...580A..40B}, which is a further argument that our blue population corresponds to a distinct thick disk. The intermediate population clearly stands out in most of these relations, even when considering the measurement uncertainties. We will discuss these stars in more detail in \sect{intermediate}.

Whether the thick disk has a distinct formation history is still much debated in the literature. A potentially separate thick disk was first recognised in vertical density profiles presented by \cite{1983MNRAS.202.1025G} and recently supported by the distribution of \sn\ stars in the [$\alpha$/Fe]-[Fe/H] plane \citep[e.g.][]{2012A&A...545A..32A, bensby+14, hayden+15} or [C/N]-[Fe/H] plane \citep{2015MNRAS.453.1855M}. In the classical scenario of Milky Way formation, the thick disk formed very early on with the bulge from the primordial protogalactic gas cloud that collapsed more than $12\,$Gyr ago \citep[see e.g.][for recent overview]{2016arXiv160801698K}. It is currently believed that thin and thick disks are distinct due to different formation processes of the disk in the past, although some authors have presented models where the thick disk arises from radial migration and/or the heating and subsequent flaring of old stars originally formed on near-circular orbits in the thin disk \citep[e.g.][]{schonrich+09,loebman+11}. \cite{aumer+16} also perform a series of controlled N-body simulations of growing disk galaxies within live dark matter haloes to steady the effect of combined spiral and giant molecular cloud heating on the disks. In these models the  outward-migrating populations are not hot enough vertically to create thick disks.  Other chemodynamical models and cosmological simulations \citep{1997ApJ...477..765C, 2011ApJ...729...16K, 2013A&A...558A...9M, minchev+15} consider further infall of gas via e.g. mergers which agree with many observables, such as the height of the disk and the metallicity and abundance gradients as a function of radius. Unfortunately,  the mechanisms of mergers for instance are still poorly understood.


Our results support a separation between thin and thick disk populations represented by two independent branches of stars evolving at different rates, suggesting that these populations have different formation histories.  

\subsection{The contribution of dynamical processes} 

Dynamical processes such as disk heating and radial migration bring in stars to the \sn\ that have different birth radii. Therefore metal-richer stars born in the inner Galaxy and metal-poorer stars born in the outer Galaxy can be brought into the \sn. The relation between branch length and stellar age in \fig{migration} for our small sample of stars however suggests that chemical enrichment is the primary driver for changes in chemical distances in the stellar populations, even in the component identified as the thin disk, where dynamical processes are thought to be more prevalent. This may be because disk-heated and radially migrated stars that came from very different parts of the Galaxy end up in a different stellar population (e.g. the intermediate stars) or not allocated to any stellar population (the undetermined stars). However there is also a scatter in the branch length-stellar age relation for thin-disk stars, which could arise from disk-heated and radially migrated stars whose birth origin is not too far from the \sn. They could have been formed from similarly chemically enriched gas at the same stellar age.

\fig{pops} shows that the contribution to the scatter in the BLA relation (\fig{migration}) for the thin-disk stars arises primarily from scatter in the [Fe/H], and to some extent scatter in other elements of the lower panel of that Figure. This suggests that some elements are more vulnerable to radial gas flows than others, and therefore are good tracers of dynamical processes. On the other hand, the elements of the upper panels of \fig{pops} have a well-defined trend suggesting that they are good tracers for studying chemical enrichment history. 

\subsection{What is the origin of the intermediate population? }\label{intermediate}

The intermediate population is significantly distinct in the chemical plots of \fig{pops} except in the case of Fe, for which three of them are slightly metal-poor and the fourth is metal-richer. They cluster in other elemental abundances and age, but have lower levels of the $\alpha$ elements (Mg, Si, Ca, S, and O) and Al, Na, Cu, Sc, and Zn than stars from the thin or thick disks with the same \ymg. The stars have a spread in actions and lie on the same trend in age and eccentricity as the other stars in the sample, i.e. they are part of the same dynamical system as the other stars.  Stars with intermediate $\alpha$ abundances have already been noted in the past \citep{1993A&AS..102..603E} and recently discussions of their nature have become more active. They might be related to the  high-$\alpha$ metal-rich stars discussed in \cite{2012A&A...545A..32A} as they were shown to be older and more $\alpha-$enhanced than the typical thin-disk stars.   Intermediate$-\alpha$ stars have also been recently discussed in \cite{2016arXiv160907821A}, who found them to display intermediate dynamical behaviour between thin and thick disk, as in this study. Whether such stars are part of the thin or the thick disk is still not known. 

\begin{figure}
\includegraphics[scale=0.48]{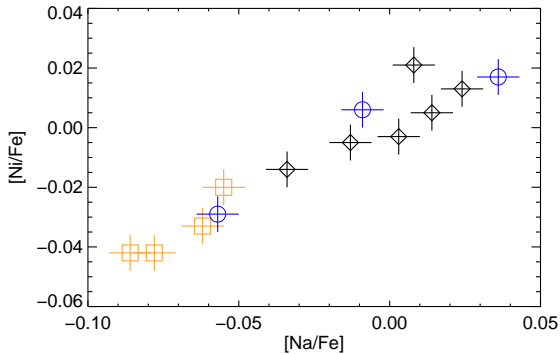}
\caption{Sodium- versus nickel-to-iron abundance ratios for the stars older and kinematically hotter than the thin-disk population. }
\label{nani}
\end{figure}

The intermediate stars could have been born at a different radius in the Milky Way, and brought into the \sn\ by radial migration or disk heating. However we would not expect such stars to cluster so strongly in abundance and age. Their chemistry suggests that they were born from a gas cloud that suffered from a different chemical enrichment history than the rest of the stars in our sample. A accretion of a smaller system in the past is a possibility. The shallower gravitational potential in smaller systems is less efficient at retaining the gas of stellar explosions that occur when massive stars die. Therefore elements such as Na, Ni, Cu, Zn and $\alpha$-elements, which are mostly produced in massive stars \citep[see][for a review of the production sites of these elements]{2013ARA&A..51..457N}, tend to be lower in these systems \citep{2009ARA&A..47..371T, 2010A&A...511L..10N, bensby+14}. Furthermore, as extensively discussed in these papers, in the [Na/Fe]-[Ni/Fe] diagram, accreted stars appear systematically at the lower end.  In \fig{nani} we show these abundance ratios for our sample of stars. Here, following e.g. \cite{bensby+14}, we omit the thin-disk (red) stars as they correspond to the kinematically  cold population. We can see that the intermediate population has lower [Na/Fe] and [Ni/Fe] abundance ratios than the thick disk stars, consistent with these stars being accreted.   

Dwarf galaxies as potential progenitors for these stars are probably ruled out as those in the Local Group tend to have much lower metallicities than solar. Only Sagittarius achieves solar metallicity \citep[e.g.][]{2009ARA&A..47..371T}. If a merger is responsible, it must have occurred around the time the stars were born, as the age-eccentricity plot shows that they have become part of the Milky Way dynamical system. Furthermore, since the stars have solar metallicities, the progenitor must have been significantly more massive than typical dwarf galaxies.    

Studies focused on the density profiles of metal-poor halo stars, such as that one of \cite{2013ApJ...763..113D}, suggest that the Milky Way broken profile can be associated with an early and massive accretion event. Several works have found signatures of past accretion events in the Milky Way by assessing clustering in age and abundance space \citep[e.g.][]{helmi+06,bensby+14}. However, \cite{ruchti+15} searched for evidence of an accreted disk component in a sample of almost 5000 stars in the Gaia-ESO spectroscopic survey with no success. This may be because they initially searched in phase space, in which these stars may already be dispersed, and they only used [Mg/Fe] abundance ratios to search for chemically distinct stars, or because there is simply none of such stars in the \sn. Since evidence of an early accreted material of solar-metallicity stars in the \sn\ have not been found \citep[see][and references therein]{ruchti+15}, it seems difficult to ensure that the intermediate population corresponds indeed to such stars given the relatively large number of them we find in our small sample.     A more detailed dynamical analysis of these interesting stars would be required to understand their origin. 

\subsection{How does a phylogenetic approach compare to other methods?}\label{other_methods}

The use of chemical patterns of stars as a method for identifying separate populations that have been either born at different radii, different Galactic components, or even extragalactic systems, already forms the basis of chemical tagging studies \citep{2002ARA&A..40..487F, 2010ApJ...713..166B}. In its simplest form, populations of \sn\ stars are { separated} in the  $[\alpha/\mathrm{Fe}]$-$[\mathrm{Fe/H}]$ diagram \citep{2014A&A...572A..33M, 2015MNRAS.453..758H} or Toomre diagram \citep{2003A&A...410..527B, 2015MNRAS.454..649H} to identify { different populations, in particular the thin and thick disks}. 
Slicing populations according to some criterion, as in the earlier studies can induce biases. They can be of a kinematic or chemical nature. { For example, it is not fully established yet that the thick disk defined as the ``high$-\alpha$/Fe" disk is the same as the kinematically hot disk or the old disk selected from [C/N] abundances \citep{2015MNRAS.453.1855M} for example.  This leads to the problem that slicing populations like this makes it very difficult to study the interface} between the thin-disk and the thick-disk  populations, especially at solar metallicities  \citep[see discussion in e.g.][]{2012A&A...545A..32A, 2016MNRAS.461.4246W} { where the $\alpha$/Fe ratios of both populations are very similar}. 

{ Furthermore, there is still no consensus to what is the best definition of the ``high$-\alpha/$Fe"  stars. One reason is  that  in} the $[\alpha/\mathrm{Fe}]$ - $[\mathrm{Fe/H}]$ diagram, there is an `intermediate-$\alpha$' population lying between the thin and thick-disk populations \citep{1993A&AS..102..603E}, probably related to our orange population.  This population is either split between the thin and thick disks \citep{2015A&A...582A.122K} or simply rejected from the analyses  \citep{2015MNRAS.453..758H, 2015MNRAS.453.1855M}. This becomes especially { uncertain} at solar metallicities, where the $[\alpha/\mathrm{Fe}]$ of the two populations overlap \citep[see discussions in e.g. ][]{2014ApJ...796...38N}. 

{ Last but not least, survey selection function plays a crucial role in this matter.  For example, it is not clear what happens to  the high $\alpha$/Fe disk as we  approach the Galactic Centre. It may not be the same as the high velocity dispersion
disk of the \sn. Cuts to define populations might be useful locally only. Survey selection function also affects the topology of trees, without a well-defined selection function of our dataset we can not give a global picture of how the evolution of the populations found as branches in the local tree takes place.  }
 
With the advent of new surveys, millions of high-resolution spectra are now available for a number of elements that can be used in the hope to define a unique chemical sequence. A particular challenge with an increasing number of chemical abundances is how best to combine them for identifying separate populations. \cite{ting+12,2015A&A...577A..47B} apply a Principal Component Analysis (PCA), which transforms a set of variables that are possibly correlated into a smaller set of linearly uncorrelated variables called principal components. \cite{2013MNRAS.428.2321M} combine weighted absolute distances in a large number of chemical elements between two stars into a single metric and use this as the basis for deriving a function that describes the probability that two stars of common evolutionary origin. \cite{hogg+16} identify overdensities in a 15-dimensional chemical-abundance space provided by APOGEE data using the k-means algorithm, which is a method for clustering points in a high-dimensional space. { This algorithm has also been applied as a basis in the chemical tagging work of \citep{2010ApJ...713..166B} but with less chemical elements and later on confirmed with simulations in \cite{2014Natur.513..523F}.} 
While PCA assumes that the variables are linearly correlated, the number of stellar populations needs to be defined a priori for the k-means clustering algorithm. 

A phylogenetic method offers a multivariate approach that does not assume linearity, does not need the number of stellar populations to be specified beforehand, and does not need to artificially slice populations. The implementation presented here with the NJ method improves on traditional clustering metrics, by normalising the chemical distance by the degree to which the stars vary from all other stars. Phylogeny also offers a very simple visualisation of separate stellar populations that additionally orders the stars within them according to chemical distance. This can then be used to compare chemical enrichment rates between stellar populations and obtain an insight into the importance of dynamical processes in shaping the distribution of chemical elements in samples of stars. Trees allow us to test hypotheses about both history and process and therefore their analysis goes much beyond being an efficient clustering algorithm like other chemical tagging techniques.

\subsection{Uncertainties in the analysis}
Perhaps the largest uncertainty in our analysis is the small sample size. The resulting consequence is that some connections between stars may have been missed, thus not allowing a full recovery of the chemical enrichment history.

In addition, we have only explored a single method for constructing the phylogenetic tree (NJ, see \sect{method}). Although we have assessed the robustness of applying this particular method, there are several other methods in the literature that may result in different groupings and branch lengths, and therefore a different phylogeny. There are other ways of constructing the distance matrix and translating this into a tree. There are also methods that do not rely on constructing a distance matrix. Particular types of evolutionary events can be penalized in the construction of the tree cost, and then an attempt is made to locate the tree with the smallest total cost. The maximum likelihood and Bayesian approaches assign probabilities to particular possible phylogenetic trees. The method is broadly similar to the maximum-parsimony method, but allows both varying rates of evolution as well as different probabilities of particular events. While this method is computationally expensive and therefore not easy to use for large datasets, it remains to be tested in future applications within Galactic archaeology.  

As with chemical tagging studies, the method presented here relies on the uniqueness of stellar DNA. We ascertain that as long as enough elements are used in the analysis, this should be the case \citep{hogg+16}. We need to be aware however that our sample needs to be chosen such that the chemical distances are reflecting differences in chemical evolution and not systematic differences due to internal processes happening in stars, such as atomic diffusion \citep[e.g.][]{2013A&A...555A..31G}, pollution due to binary companions \citep{1980ApJ...238L..35M} { or even enrichment due to accretion of gas from the ISM \citep{2016arXiv161202832S}.} Therefore it is important to have a sample of stars with the same spectral class \citep[see e.g. extensive discussion in][]{2015A&A...577A..47B, 2015A&A...582A..81J, 2016arXiv161205013J}. 

\section{Summary and Conclusions}\label{conclusions}

We demonstrate the potential for the use of a phylogenetic method in visualising and analysing the chemical evolution of \sn\ stars, using abundances of 17 chemical elements for the 22 solar twins of \cite{Nissen15,Nissen16} as a proxy for DNA. The chemical abundances were used to create a matrix of the chemical distances between pair of stars. This matrix was input into a software developed for molecular biology to create a phylogenetic tree. Despite the small size of the sample, we believe the method successfully recovered stellar populations with distinct chemical enrichment histories and produced a succinct visualisation of the stars in a multi-dimensional chemical space.  

A comparison of the order of the stars along tree branches with stellar ages confirmed that the order generally traces the direction of chemical enrichment, thus allowing an estimate of the mean chemical enrichment rate to be made. Chemical enrichment has a mean rate of $<\dot{\Sigma}_{XFe}> = 0.168\,\mathrm{dex} \,\mathrm{Gyr}^{-1}$ in the thin disk, $<\dot{\Sigma}_{XFe}> = 0.657\,\mathrm{dex} \,\mathrm{Gyr}^{-1}$ in the thick disk, and $<\dot{\Sigma}_{XFe}> = 0.269\,\mathrm{dex} \,\mathrm{Gyr}^{-1}$ in the intermediate population of stars. Our analysis thus confirms, in a purely empirical way, that the star formation rate in the thick disk is much faster than in the thin disk. 

In addition to confirming a likely separate chemical enrichment history in the thin disk compared to the thick disk, we also find a separate population, intermediate in age and eccentricities but distinct and clustered in several abundance ratios. Due to its old age and low abundances of $\alpha-$ and other iron-peak elements with respect to coeval stars in the thin and thick-disk populations, we surmise that these stars could have arrived via a major merger at the early stages of the Milky Way formation. We however could not rule out the possibility that these stars might be the youngest tail of the thick disk or the oldest tail of the thin disk, as they belonged to similar trends in some abundance ratios and kinematical properties. The fact  they emerge as an independent branch in our tree could be either due to a truly different nature or due to selection effects. We conclude that a more detailed dynamical study of such stars and a larger sample of old solar-metallicity stars is necessary. Future work will also benefit greatly from tests on simulated stars in the \sn\ in order to better understand how radially migrated and disk heated stars appear in the phylogenetic tree.  

In biology it is commonly said that to study evolution, one essentially analyses trees. Galactic archaeology should be no different, especially now, during its golden ages. Thanks to Gaia and its complementary spectroscopic and asteroseismic surveys, we are quickly getting chemical abundances of millions of stars which can be complemented with accurate astrometry and ages. These rich datasets are on the verge of  putting us closer to finding the one  tree that connects all stars in the Milky Way.

\subsubsection*{Acknowledgments} 
We are deeply grateful to K.M.~Reinhart, G. Weiss-Sussex and B. Burgwinkle for organising the research event  in King's College Cambridge that inspired this work.  We also thank P. E. Nissen for useful comments about this manuscript.  P.J. acknowledges C. Worley, T. Masseron, T. M\"adler and G. Gilmore and P.D.  thanks D. Labonte, J. Binney and the Oxford Galactic Dynamics group for discussions on the topic. We acknowledge the positive feedback from the referee report. The research leading to these results has received funding from the European Research Council under the European Union's Seventh Framework Programme (FP7/2007-2013)/ERC grant agreement no.s \ 320360 and \ 321067, as well as King's College Cambridge CRA programme.

\bibliography{refs_tree}

\begin{appendix}

\section{Astrometric and kinematic data}\label{ap1}

\begin{figure}
 {\hspace{-0.5cm}}
\includegraphics[scale=0.45]{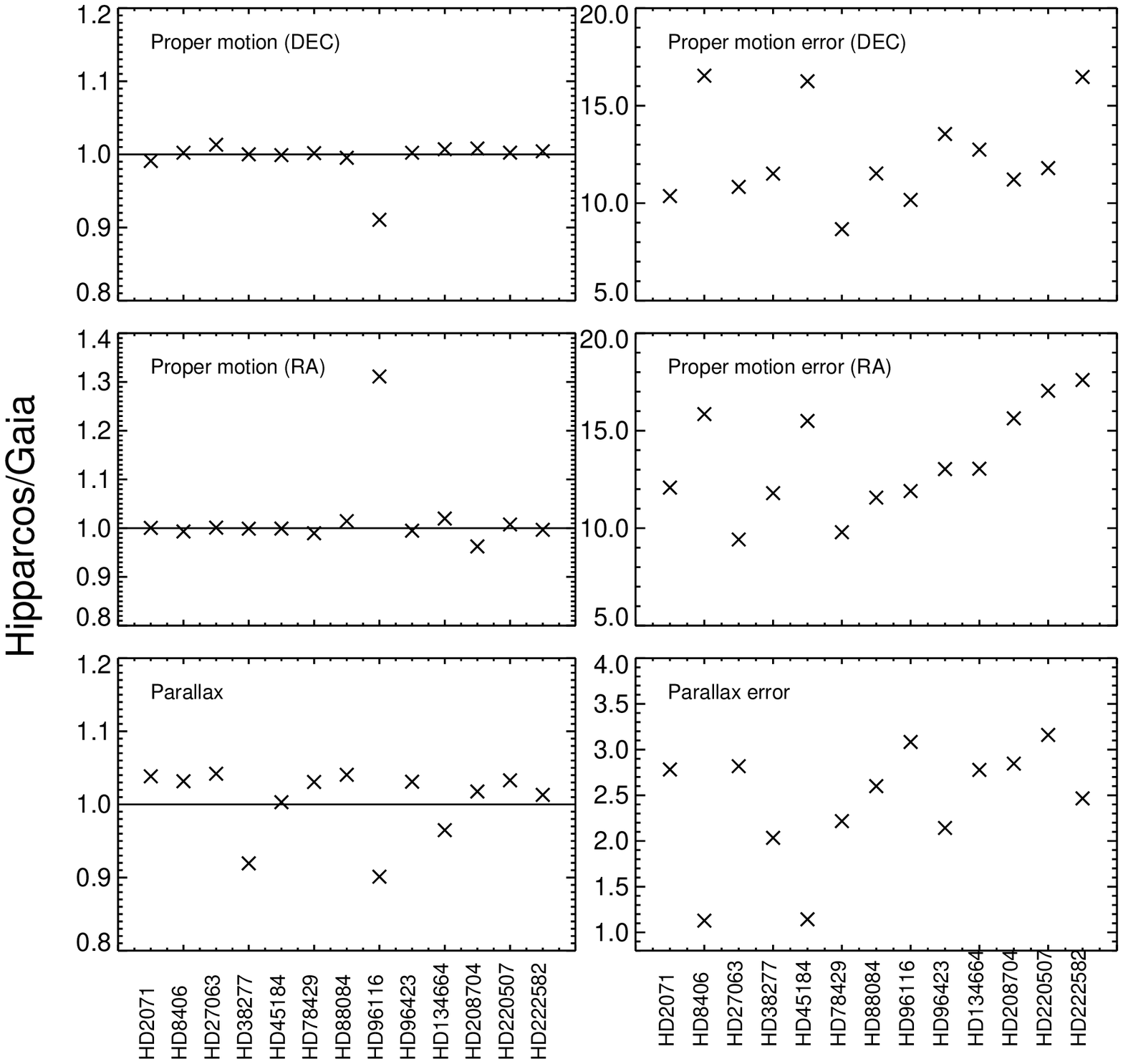}
\caption{Astrometric data of Hipparcos and Gaia compared for our sample of stars. }
\label{gaiahip}
\end{figure}

In \tab{astrometry} we list the coordinates, parallaxes and proper motions of our sample of stars from the reduction of the Hipparcos data by \cite{2007A&A...474..653V}. Since these stars have been observed with HARPS for the purpose of radial velocity monitoring to detect planets, there are several RV measurements for them available. The RV value listed in the table represents the mean of 3 measurements taken at different epochs.

\begin{table*}
\begin{center}
\input{astrometry_H.tex}
\caption{Hipparcos astrometry of the solar twins used in this work. We indicate the name of the star, equatorial sky positions, $V$-band apparent magnitudes, parallaxes ($\varpi$), proper motions (PM), the radial velocities (RV). }
\label{astrometry}
\end{center}
\end{table*}%

\subsection{Hipparcos and Gaia}
While our analysis was being performed, the first data release of Gaia became public \citep{2016arXiv160904172G}. Here we compare the astrometric data of both missions in \fig{gaiahip}.  In the left-hand panels we show the ratio of the proper motions and parallaxes of Gaia and Hipparcos in separate panels, and each star is indicated in the x-axis. We can see that the differences between both are usually within 10\%, except few cases such as HD~96116, whose proper motion in RA has a difference of 30\%. The right-hand panels show the ratio of the errors of these quantities in separate panels. Here we see how Gaia data is on general 3 times more accurate in the derived parallaxes, and up to 20 times more accurate in proper motions. 

While this comparison would suggest to use Gaia data instead of Hipparcos, we caution here that we do not have Gaia astrometry for all the stars in our sample. Since we have solar twins only, we could have applied the twin method \citep{2015MNRAS.453.1428J} to determine parallaxes of the resting stars not contained in Gaia. However, in our analysis we require proper motions which are best measured by an astrometric mission. Therefore, to keep our analysis homogeneous and complete, we decided to use the Hipparcos astrometry for our analysis of kinematics of stars. 
 As we did not find anything unusual from the kinematic behaviour of our stars, we believe that reducing the dataset to consider Gaia data has not a favourable impact nor changes the conclusions of our work.

Using the data of \tab{astrometry}, that is, the astrometry of Hipparcos and the RVs from HARPS, we calculated the actions and the total velocity of the stars as described in \sect{agesdynprop}. These values are listed in \tab{actions_ages}, together with the ages as derived by \cite{Nissen16} for completeness.

\begin{table*}
\begin{center}
\input{actions_H.tex}
\caption{Galactic coordinates, total velocities, actions and ages of the stars used in our sample. Ages are taken from \citet{Nissen16}, while other properties are determined from the astrometry of \tab{astrometry} as described in \sect{agesdynprop}.}
\label{actions_ages}
\end{center}
\end{table*}%

\end{appendix}

\end{document}

%% file: astrometry_H.tex
\begin{tabular}{|l|r|r|r|r|r|r|r|r|}
\hline
  \multicolumn{1}{|c|}{Star} &
  \multicolumn{1}{c|}{RA } &
  \multicolumn{1}{c|}{DEC } &
  \multicolumn{1}{c|}{$\varpi$ } &
  \multicolumn{1}{c|}{$\sigma \varpi$ } &
  \multicolumn{1}{c|}{V } &
  \multicolumn{1}{c|}{pm(RA)} &
  \multicolumn{1}{c|}{pm(DEC) } &
  \multicolumn{1}{c|}{RV } \\
 & [degrees] & [degrees] & [mas] & [mas] & [mag] & [mas/yr] & [mas/yr] & [km/s] \\ 
 \hline
  HD2071 & 0.1077987647 & -0.9421973172 & 36.72 & 0.64 & 7.27 & 210.72 & -27.9 & 6.6823\\
  HD8406 & 0.3622254703 & -0.2877530204 & 27.38 & 0.79 & 7.92 & -136.95 & -176.35 & -7.6766\\
  HD20782 & 0.8729078992 & -0.5035957911 & 28.15 & 0.62 & 7.36 & 349.33 & -65.92 & 39.9666\\
  HD27063 & 1.1192125265 & -0.0099900057 & 24.09 & 0.62 & 8.07 & -61.2 & -174.37 & -9.5868\\
  HD28471 & 1.1569520864 & -1.1184218496 & 23.48 & 0.52 & 7.89 & -61.21 & 321.65 & 54.8322\\
  HD38277 & 1.5022599908 & -0.1748188646 & 24.42 & 0.59 & 7.11 & 65.86 & -143.64 & 32.7339\\
  HD45184 & 1.6787150866 & -0.5023025992 & 45.7 & 0.4 & 6.37 & -164.99 & -121.77 & -3.7577\\
  HD45289 & 1.6772913373 & -0.7478632512 & 35.81 & 0.32 & 6.67 & -77.14 & 777.98 & 56.4638\\
  HD71334 & 2.207072235 & -0.5223749846 & 26.64 & 0.78 & 7.81 & 139.04 & -292.03 & 17.3836\\
  HD78429 & 2.3851954548 & -0.7590858065 & 26.83 & 0.51 & 7.31 & 48.18 & 179.49 & 65.1012\\
  HD88084 & 2.6578675783 & -0.2704215986 & 29.01 & 0.65 & 7.52 & -93.59 & -196.69 & -23.5602\\
  HD92719 & 2.8022119328 & -0.2406336934 & 41.97 & 0.47 & 6.79 & 235.35 & -172.56 & -17.8906\\
  HD96116 & 2.8982366461 & -1.0081901354 & 16.96 & 0.74 & 8.65 & 31.49 & -35.4 & 31.3132\\
  HD96423 & 2.9074038637 & -0.7744568097 & 31.87 & 0.6 & 7.23 & 87.3 & -86.9 & 54.8541\\
  HD134664 & 3.9800988033 & -0.5390630337 & 23.36 & 0.75 & 7.76 & 99.89 & -105.82 & 7.6595\\
  HD146233 & 4.2569404686 & -0.1460532782 & 71.94 & 0.37 & 5.49 & 230.77 & -495.53 & 11.8365\\
  HD183658 & 5.1089279481 & -0.113692144 & 31.93 & 0.6 & 7.27 & -142.51 & -141.03 & 58.286\\
  HD208704 & 5.7526274579 & -0.2210426045 & 29.93 & 0.74 & 7.16 & 32.03 & 62.24 & 3.9277\\
  HD210918 & 5.8234517182 & -0.7222127624 & 45.35 & 0.37 & 6.23 & 571.11 & -789.84 & -19.0537\\
  HD220507 & 6.1291695816 & -0.9198146045 & 23.79 & 0.79 & 7.59 & -17.24 & -157.93 & 23.2956\\
  HD222582 & 6.2040357329 & -0.1044664586 & 23.94 & 0.74 & 7.68 & -144.88 & -111.93 & 12.0876\\
\hline\end{tabular}

%% file: actions_H.tex
\begin{tabular}{|l|r|r|r|r|r|r|r|r|r|}
\hline
  \multicolumn{1}{|c|}{Star} &
  \multicolumn{1}{c|}{$R$} &
  \multicolumn{1}{c|}{$z$} &
  \multicolumn{1}{c|}{$U$}&
  \multicolumn{1}{c|}{$V$} &
  \multicolumn{1}{c|}{$W$} &
  \multicolumn{1}{c|}{$J_r$}&
  \multicolumn{1}{c|}{$J_z$} &
  \multicolumn{1}{c|}{$Lz$}&
  \multicolumn{1}{c|}{Age} \\
  & [kpc] & [kpc] & [km/s] & [km/s] & [km/s] & [kpc km/s] & [kpc km/s] & [kpc km/s] & [Gyr] \\
  \hline
  HD2071 &8.29172&0.00113&-19.95267&-18.63223&-7.14419&14.91899&0.00585&-1920.49109& 3.5\\
  HD8406 &8.30762&-0.00035&37.76446&-11.09508&-1.11882&6.75647&0.21467&-1986.76341& 4.1\\
  HD20782 &8.31363&0.00163&-37.38056&-61.41219&-2.65999&78.45743&0.12547&-1570.39096& 8.1\\
  HD27063 &8.33342&0.00096&28.59609&-16.04689&-18.43578&3.46062&0.70879&-1951.65914& 2.8\\
  HD28471 &8.29683&0.00392&-53.89275&-25.00513&-62.07675&55.17953&28.08048&-1870.88032& 7.3\\
  HD38277 &8.33184&0.00262&-7.98824&-42.68783&-11.25881&27.42507&0.07784&-1730.0909& 7.8\\
  HD45184 &8.31146&0.00209&10.02202&5.31122&-18.37438&5.65876&0.68668&-2124.07984& 2.7\\
  HD45289* &8.30851&0.00293&-115.57143&-22.89469&-3.77961&205.68372&0.05632&-1892.0283& 9.4\\
  HD71334 &8.3128&0.00423&47.27097&-36.36525&-8.19526&27.75369&0.01968&-1777.03046& 8.8\\
  HD78429 &8.3028&0.00447&-22.35954&-62.31564&30.54011&62.0256&11.90181&-1561.62639& 8.3\\
  HD88084 &8.30748&0.00341&10.91872&-0.78358&-41.25848&2.60847&8.91765&-2072.37971& 5.9\\
  HD92719 &8.30281&0.00222&34.73916&9.3428&-10.60875&14.70023&0.09766&-2154.85832& 2.5\\
  HD96116 &8.28099&0.00673&21.61917&-25.8706&-4.33059&6.63941&0.05597&-1857.59697& 0.7\\
  HD96423 &8.29279&0.00355&29.72398&-49.04196&7.4871&33.82754&1.34169&-1667.99685& 7.3\\
  HD134664 &8.2642&0.00198&18.77267&-3.22958&-23.82393&1.42554&1.73021&-2041.25362& 2.4\\
  HD146233 &8.28791&-0.00012&27.17963&-14.42555&-22.1416&2.48326&1.33952&-1954.43792& 4.0\\
  HD183658 &8.27389&-0.00194&65.03352&6.90413&-2.41432&39.79151&0.11846&-2128.34777& 5.2\\
  HD208704 &8.28359&-0.0019&-5.84723&9.76923&-2.99267&14.57429&0.09451&-2153.53487& 7.4\\
  HD210918* &8.28735&0.00004&-47.58932&-91.6203&-8.57114&155.99919&0.00903&-1314.04448& 9.1\\
  HD220507* &8.28179&0.0013&20.30581&-32.88105&-7.00157&11.95045&0.02781&-1799.71119& 9.8\\
  HD222582 &8.29731&-0.00225&36.52928&-0.67894&-11.23786&7.96257&0.10894&-2071.02691& 7.0\\
  Sun &8.3&0.014&0&0&0&4.68053&0.28028&-2076.992& 4.5\\
\hline\end{tabular}

%% file: NissenTree_v6_submP.bbl
\begin{thebibliography}{77}
\expandafter\ifx\csname natexlab\endcsname\relax\def\natexlab#1{#1}\fi

\bibitem[{{Adibekyan} {et~al}\mbox{.}(2012){Adibekyan}, {Sousa}, {Santos},
  {Delgado Mena}, {Gonz{\'a}lez Hern{\'a}ndez}, {Israelian}, {Mayor}, \&
  {Khachatryan}}]{2012A&A...545A..32A}
{Adibekyan} V.~Z., {Sousa} S.~G., {Santos} N.~C., {Delgado Mena} E.,
  {Gonz{\'a}lez Hern{\'a}ndez} J.~I., {Israelian} G., {Mayor} M., {Khachatryan}
  G., 2012, \aap, 545, A32

\bibitem[{{Allende Prieto}, {Kawata} \& {Cropper}(2016){Allende Prieto},
  {Kawata}, \& {Cropper}}]{2016arXiv160907821A}
{Allende Prieto} C., {Kawata} D., {Cropper} M., 2016, ArXiv e-prints

\bibitem[{{Aumer}, {Binney} \& {Sch{\"o}nrich}(2016){Aumer}, {Binney}, \&
  {Sch{\"o}nrich}}]{aumer+16}
{Aumer} M., {Binney} J., {Sch{\"o}nrich} R., 2016, \mnras, 459, 3326

\bibitem[{{Barbuy} {et~al}\mbox{.}(2015){Barbuy}, {Fria{\c c}a}, {da Silveira},
  {Hill}, {Zoccali}, {Minniti}, {Renzini}, {Ortolani}, \&
  {G{\'o}mez}}]{2015A&A...580A..40B}
{Barbuy} B. {et~al.}, 2015, \aap, 580, A40

\bibitem[{{Bensby}, {Feltzing} \& {Lundstr{\"o}m}(2003){Bensby}, {Feltzing}, \&
  {Lundstr{\"o}m}}]{2003A&A...410..527B}
{Bensby} T., {Feltzing} S., {Lundstr{\"o}m} I., 2003, \aap, 410, 527

\bibitem[{{Bensby}, {Feltzing} \& {Oey}(2014){Bensby}, {Feltzing}, \&
  {Oey}}]{bensby+14}
{Bensby} T., {Feltzing} S., {Oey} M.~S., 2014, \aap, 562, A71

\bibitem[{{Blanco-Cuaresma} {et~al}\mbox{.}(2015){Blanco-Cuaresma}, {Soubiran},
  {Heiter}, {Asplund}, {Carraro}, {Costado}, {Feltzing},
  {Gonz{\'a}lez-Hern{\'a}ndez}, {Jim{\'e}nez-Esteban}, {Korn}, {Marino},
  {Montes}, {San Roman}, {Tabernero}, \& {Tautvai{\v
  s}ien{\.e}}}]{2015A&A...577A..47B}
{Blanco-Cuaresma} S. {et~al.}, 2015, \aap, 577, A47

\bibitem[{{Bland-Hawthorn}, {Krumholz} \& {Freeman}(2010){Bland-Hawthorn},
  {Krumholz}, \& {Freeman}}]{2010ApJ...713..166B}
{Bland-Hawthorn} J., {Krumholz} M.~R., {Freeman} K., 2010, \apj, 713, 166

\bibitem[{{Chiappini}, {Matteucci} \& {Gratton}(1997){Chiappini}, {Matteucci},
  \& {Gratton}}]{1997ApJ...477..765C}
{Chiappini} C., {Matteucci} F., {Gratton} R., 1997, \apj, 477, 765

\bibitem[{Dan \& Li(2000)}]{dan2000fundamentals}
Dan G., Li W.-H., 2000, Fundamentals of molecular evolution

\bibitem[{Darwin(1859)}]{darwin2003origin}
Darwin C., 1859, Ed. Joseph Carroll. Toronto: Broadview, 2003 edition

\bibitem[{{Datson}, {Flynn} \& {Portinari}(2014){Datson}, {Flynn}, \&
  {Portinari}}]{2014MNRAS.439.1028D}
{Datson} J., {Flynn} C., {Portinari} L., 2014, \mnras, 439, 1028

\bibitem[{{De Silva} {et~al}\mbox{.}(2015){De Silva}, {Freeman},
  {Bland-Hawthorn}, {Martell}, {de Boer}, {Asplund}, {Keller}, {Sharma},
  {Zucker}, {Zwitter}, {Anguiano}, {Bacigalupo}, {Bayliss}, {Beavis},
  {Bergemann}, {Campbell}, {Cannon}, {Carollo}, {Casagrande}, {Casey}, {Da
  Costa}, {D'Orazi}, {Dotter}, {Duong}, {Heger}, {Ireland}, {Kafle}, {Kos},
  {Lattanzio}, {Lewis}, {Lin}, {Lind}, {Munari}, {Nataf}, {O'Toole}, {Parker},
  {Reid}, {Schlesinger}, {Sheinis}, {Simpson}, {Stello}, {Ting}, {Traven},
  {Watson}, {Wittenmyer}, {Yong}, \& {{\v Z}erjal}}]{2015MNRAS.449.2604D}
{De Silva} G.~M. {et~al.}, 2015, \mnras, 449, 2604

\bibitem[{{Deason} {et~al}\mbox{.}(2013){Deason}, {Belokurov}, {Evans}, \&
  {Johnston}}]{2013ApJ...763..113D}
{Deason} A.~J., {Belokurov} V., {Evans} N.~W., {Johnston} K.~V., 2013, \apj,
  763, 113

\bibitem[{{Dehnen} \& {Binney}(1998)}]{dehnen+98}
{Dehnen} W., {Binney} J., 1998, \mnras, 294, 429

\bibitem[{{Edvardsson} {et~al}\mbox{.}(1993){Edvardsson}, {Andersen},
  {Gustafsson}, {Lambert}, {Nissen}, \& {Tomkin}}]{1993A&AS..102..603E}
{Edvardsson} B., {Andersen} J., {Gustafsson} B., {Lambert} D.~L., {Nissen}
  P.~E., {Tomkin} J., 1993, \aaps, 102, 603

\bibitem[{{Felsenstein}(1982)}]{Felsenstein}
{Felsenstein} J., 1982, The Quarterly Review of Biology, 57, 379

\bibitem[{Felsenstein(1988)}]{felsenstein1988phylogenies}
Felsenstein J., 1988, Annual review of genetics, 22, 521

\bibitem[{{Feltzing} {et~al}\mbox{.}(2016){Feltzing}, {Howes}, {McMillan}, \&
  {Stonkute}}]{2016arXiv161003852F}
{Feltzing} S., {Howes} L.~M., {McMillan} P.~J., {Stonkute} E., 2016, ArXiv
  e-prints

\bibitem[{{Feng} \& {Krumholz}(2014)}]{2014Natur.513..523F}
{Feng} Y., {Krumholz} M.~R., 2014, \nat, 513, 523

\bibitem[{{Fraix-Burnet}, {Choler} \& {Douzery}(2006){Fraix-Burnet}, {Choler},
  \& {Douzery}}]{2006A&A...455..845F}
{Fraix-Burnet} D., {Choler} P., {Douzery} E.~J.~P., 2006, \aap, 455, 845

\bibitem[{{Fraix-Burnet} \& {Davoust}(2015)}]{2015MNRAS.450.3431F}
{Fraix-Burnet} D., {Davoust} E., 2015, \mnras, 450, 3431

\bibitem[{{Fraix-Burnet}, {Davoust} \& {Charbonnel}(2009){Fraix-Burnet},
  {Davoust}, \& {Charbonnel}}]{2009MNRAS.398.1706F}
{Fraix-Burnet} D., {Davoust} E., {Charbonnel} C., 2009, \mnras, 398, 1706

\bibitem[{{Freeman} \& {Bland-Hawthorn}(2002)}]{2002ARA&A..40..487F}
{Freeman} K., {Bland-Hawthorn} J., 2002, \araa, 40, 487

\bibitem[{{Gaia Collaboration} {et~al}\mbox{.}(2016){Gaia Collaboration},
  {Brown}, {Vallenari}, {Prusti}, {de Bruijne}, {Mignard}, {Drimmel}, \&
  {co-authors}}]{2016arXiv160904172G}
{Gaia Collaboration}, {Brown} A.~G.~A., {Vallenari} A., {Prusti} T., {de
  Bruijne} J., {Mignard} F., {Drimmel} R., {co-authors} ., 2016, ArXiv e-prints

\bibitem[{{Gilmore} \& {Reid}(1983)}]{1983MNRAS.202.1025G}
{Gilmore} G., {Reid} N., 1983, \mnras, 202, 1025

\bibitem[{{Gruyters} {et~al}\mbox{.}(2013){Gruyters}, {Korn}, {Richard},
  {Grundahl}, {Collet}, {Mashonkina}, {Osorio}, \&
  {Barklem}}]{2013A&A...555A..31G}
{Gruyters} P., {Korn} A.~J., {Richard} O., {Grundahl} F., {Collet} R.,
  {Mashonkina} L.~I., {Osorio} Y., {Barklem} P.~S., 2013, \aap, 555, A31

\bibitem[{{Hattori} \& {Gilmore}(2015)}]{2015MNRAS.454..649H}
{Hattori} K., {Gilmore} G., 2015, \mnras, 454, 649

\bibitem[{{Hawkins} {et~al}\mbox{.}(2015){Hawkins}, {Jofr{\'e}}, {Masseron}, \&
  {Gilmore}}]{2015MNRAS.453..758H}
{Hawkins} K., {Jofr{\'e}} P., {Masseron} T., {Gilmore} G., 2015, \mnras, 453,
  758

\bibitem[{{Hayden} {et~al}\mbox{.}(2015){Hayden}, {Bovy}, {Holtzman},
  {Nidever}, {Bird}, {Weinberg}, {Andrews}, {Majewski}, {Allende Prieto},
  {Anders}, {Beers}, {Bizyaev}, {Chiappini}, {Cunha}, {Frinchaboy},
  {Garc{\'{\i}}a-Her{\'n}andez}, {Garc{\'{\i}}a P{\'e}rez}, {Girardi},
  {Harding}, {Hearty}, {Johnson}, {M{\'e}sz{\'a}ros}, {Minchev}, {O'Connell},
  {Pan}, {Robin}, {Schiavon}, {Schneider}, {Schultheis}, {Shetrone},
  {Skrutskie}, {Steinmetz}, {Smith}, {Wilson}, {Zamora}, \&
  {Zasowski}}]{hayden+15}
{Hayden} M.~R. {et~al.}, 2015, \apj, 808, 132

\bibitem[{{Helmi} {et~al}\mbox{.}(2006){Helmi}, {Navarro}, {Nordstr{\"o}m},
  {Holmberg}, {Abadi}, \& {Steinmetz}}]{helmi+06}
{Helmi} A., {Navarro} J.~F., {Nordstr{\"o}m} B., {Holmberg} J., {Abadi} M.~G.,
  {Steinmetz} M., 2006, \mnras, 365, 1309

\bibitem[{{Hogg} {et~al}\mbox{.}(2016){Hogg}, {Casey}, {Ness}, {Rix},
  {Foreman-Mackey}, {Hasselquist}, {Ho}, {Holtzman}, {Majewski}, {Martell},
  {Meszaros}, {NIdever}, \& {Shetrone}}]{hogg+16}
{Hogg} D.~W. {et~al.}, 2016, ArXiv e-prints

\bibitem[{{Jofr{\'e}} {et~al}\mbox{.}(2015{\natexlab{a}}){Jofr{\'e}}, {Heiter},
  {Soubiran}, {Blanco-Cuaresma}, {Masseron}, {Nordlander}, {Chemin}, {Worley},
  {Van Eck}, {Hourihane}, {Gilmore}, {Adibekyan}, {Bergemann}, {Cantat-Gaudin},
  {Delgado-Mena}, {Gonz{\'a}lez Hern{\'a}ndez}, {Guiglion}, {Lardo}, {de
  Laverny}, {Lind}, {Magrini}, {Mikolaitis}, {Montes}, {Pancino},
  {Recio-Blanco}, {Sordo}, {Sousa}, {Tabernero}, \&
  {Vallenari}}]{2015A&A...582A..81J}
{Jofr{\'e}} P. {et~al.}, 2015{\natexlab{a}}, \aap, 582, A81

\bibitem[{{Jofr\'e} {et~al}\mbox{.}(2016){Jofr\'e}, {Heiter}, {Worley},
  {Blanco-Cuaresma}, {Soubiran}, {Masseron}, {Hawkins}, {Adibekyan}, {Buder},
  {Casamiquela}, {Gilmore}, {Hourihane}, \& {Tabernero}}]{2016arXiv161205013J}
{Jofr\'e} P. {et~al.}, 2016, ArXiv e-prints

\bibitem[{{Jofr{\'e}} {et~al}\mbox{.}(2015{\natexlab{b}}){Jofr{\'e}},
  {M{\"a}dler}, {Gilmore}, {Casey}, {Soubiran}, \&
  {Worley}}]{2015MNRAS.453.1428J}
{Jofr{\'e}} P., {M{\"a}dler} T., {Gilmore} G., {Casey} A.~R., {Soubiran} C.,
  {Worley} C., 2015{\natexlab{b}}, \mnras, 453, 1428

\bibitem[{{Kawata} \& {Chiappini}(2016)}]{2016arXiv160801698K}
{Kawata} D., {Chiappini} C., 2016, ArXiv e-prints

\bibitem[{{Kobayashi} \& {Nakasato}(2011)}]{2011ApJ...729...16K}
{Kobayashi} C., {Nakasato} N., 2011, \apj, 729, 16

\bibitem[{{Kordopatis} {et~al}\mbox{.}(2015){Kordopatis}, {Wyse}, {Gilmore},
  {Recio-Blanco}, {de Laverny}, {Hill}, {Adibekyan}, {Heiter}, {Minchev},
  {Famaey}, {Bensby}, {Feltzing}, {Guiglion}, {Korn}, {Mikolaitis},
  {Schultheis}, {Vallenari}, {Bayo}, {Carraro}, {Flaccomio}, {Franciosini},
  {Hourihane}, {Jofr{\'e}}, {Koposov}, {Lardo}, {Lewis}, {Lind}, {Magrini},
  {Morbidelli}, {Pancino}, {Randich}, {Sacco}, {Worley}, \&
  {Zaggia}}]{2015A&A...582A.122K}
{Kordopatis} G. {et~al.}, 2015, \aap, 582, A122

\bibitem[{Kumar, Stecher \& Tamura(2016)Kumar, Stecher, \&
  Tamura}]{kumar2016mega7}
Kumar S., Stecher G., Tamura K., 2016, Molecular biology and evolution, msw054

\bibitem[{Lemey(2009)}]{lemey2009phylogenetic}
Lemey P., 2009, The phylogenetic handbook: a practical approach to phylogenetic
  analysis and hypothesis testing. Cambridge University Press

\bibitem[{{Loebman} {et~al}\mbox{.}(2011){Loebman}, {Ro{\v s}kar},
  {Debattista}, {Ivezi{\'c}}, {Quinn}, \& {Wadsley}}]{loebman+11}
{Loebman} S.~R., {Ro{\v s}kar} R., {Debattista} V.~P., {Ivezi{\'c}} {\v Z}.,
  {Quinn} T.~R., {Wadsley} J., 2011, \apj, 737, 8

\bibitem[{{Masseron} \& {Gilmore}(2015)}]{2015MNRAS.453.1855M}
{Masseron} T., {Gilmore} G., 2015, \mnras, 453, 1855

\bibitem[{{Matteucci}(2012)}]{mattuecci+12}
{Matteucci} F., 2012, {Chemical Evolution of Galaxies}

\bibitem[{{McClure}, {Fletcher} \& {Nemec}(1980){McClure}, {Fletcher}, \&
  {Nemec}}]{1980ApJ...238L..35M}
{McClure} R.~D., {Fletcher} J.~M., {Nemec} J.~M., 1980, \apjl, 238, L35

\bibitem[{{McWilliam} \& {Rauch}(2004)}]{mcwilliam+04}
{McWilliam} A., {Rauch} M., 2004, Origin and Evolution of the Elements

\bibitem[{{Mel{\'e}ndez}, {Dodds-Eden} \& {Robles}(2006){Mel{\'e}ndez},
  {Dodds-Eden}, \& {Robles}}]{2006ApJ...641L.133M}
{Mel{\'e}ndez} J., {Dodds-Eden} K., {Robles} J.~A., 2006, \apjl, 641, L133

\bibitem[{{Mikolaitis} {et~al}\mbox{.}(2014){Mikolaitis}, {Hill},
  {Recio-Blanco}, {de Laverny}, {Allende Prieto}, {Kordopatis}, {Tautvai{\v
  s}iene}, {Romano}, {Gilmore}, {Randich}, {Feltzing}, {Micela}, {Vallenari},
  {Alfaro}, {Bensby}, {Bragaglia}, {Flaccomio}, {Lanzafame}, {Pancino},
  {Smiljanic}, {Bergemann}, {Carraro}, {Costado}, {Damiani}, {Hourihane},
  {Jofr{\'e}}, {Lardo}, {Magrini}, {Maiorca}, {Morbidelli}, {Sbordone},
  {Sousa}, {Worley}, \& {Zaggia}}]{2014A&A...572A..33M}
{Mikolaitis} {\v S}. {et~al.}, 2014, \aap, 572, A33

\bibitem[{{Minchev}, {Chiappini} \& {Martig}(2013){Minchev}, {Chiappini}, \&
  {Martig}}]{2013A&A...558A...9M}
{Minchev} I., {Chiappini} C., {Martig} M., 2013, \aap, 558, A9

\bibitem[{{Minchev} \& {Famaey}(2010)}]{2010ApJ...722..112M}
{Minchev} I., {Famaey} B., 2010, \apj, 722, 112

\bibitem[{{Minchev} {et~al}\mbox{.}(2015){Minchev}, {Martig}, {Streich},
  {Scannapieco}, {de Jong}, \& {Steinmetz}}]{minchev+15}
{Minchev} I., {Martig} M., {Streich} D., {Scannapieco} C., {de Jong} R.~S.,
  {Steinmetz} M., 2015, \apjl, 804, L9

\bibitem[{{Mitschang} {et~al}\mbox{.}(2013){Mitschang}, {De Silva}, {Sharma},
  \& {Zucker}}]{2013MNRAS.428.2321M}
{Mitschang} A.~W., {De Silva} G., {Sharma} S., {Zucker} D.~B., 2013, \mnras,
  428, 2321

\bibitem[{{Nidever} {et~al}\mbox{.}(2014){Nidever}, {Bovy}, {Bird}, {Andrews},
  {Hayden}, {Holtzman}, {Majewski}, {Smith}, {Robin}, {Garc{\'{\i}}a
  P{\'e}rez}, {Cunha}, {Allende Prieto}, {Zasowski}, {Schiavon}, {Johnson},
  {Weinberg}, {Feuillet}, {Schneider}, {Shetrone}, {Sobeck},
  {Garc{\'{\i}}a-Hern{\'a}ndez}, {Zamora}, {Rix}, {Beers}, {Wilson},
  {O'Connell}, {Minchev}, {Chiappini}, {Anders}, {Bizyaev}, {Brewington},
  {Ebelke}, {Frinchaboy}, {Ge}, {Kinemuchi}, {Malanushenko}, {Malanushenko},
  {Marchante}, {M{\'e}sz{\'a}ros}, {Oravetz}, {Pan}, {Simmons}, \&
  {Skrutskie}}]{2014ApJ...796...38N}
{Nidever} D.~L. {et~al.}, 2014, \apj, 796, 38

\bibitem[{{Nissen}(2015)}]{Nissen15}
{Nissen} P.~E., 2015, \aap, 579, A52

\bibitem[{{Nissen}(2016)}]{Nissen16}
{Nissen} P.~E., 2016, \aap, 593, A65

\bibitem[{{Nissen} \& {Schuster}(2010)}]{2010A&A...511L..10N}
{Nissen} P.~E., {Schuster} W.~J., 2010, \aap, 511, L10

\bibitem[{{Nomoto}, {Kobayashi} \& {Tominaga}(2013){Nomoto}, {Kobayashi}, \&
  {Tominaga}}]{2013ARA&A..51..457N}
{Nomoto} K., {Kobayashi} C., {Tominaga} N., 2013, \araa, 51, 457

\bibitem[{{Piffl} {et~al}\mbox{.}(2014){Piffl}, {Binney}, {McMillan},
  {Steinmetz}, {Helmi}, {Wyse}, {Bienaym{\'e}}, {Bland-Hawthorn}, \&
  {al}}]{piffl+14}
{Piffl} T. {et~al.}, 2014, \mnras, 445, 3133

\bibitem[{Ren, Ishida \& Akiyama(2013)Ren, Ishida, \&
  Akiyama}]{ren2013assessing}
Ren A., Ishida T., Akiyama Y., 2013, Molecular phylogenetics and evolution, 67,
  429

\bibitem[{Ridley(1986)}]{ridley1986evolution}
Ridley M., 1986

\bibitem[{Ropiquet, Li \& Hassanin(2009)Ropiquet, Li, \&
  Hassanin}]{ropiquet2009supertri}
Ropiquet A., Li B., Hassanin A., 2009, Comptes rendus biologies, 332, 832

\bibitem[{{Ruchti} {et~al}\mbox{.}(2015){Ruchti}, {Read}, {Feltzing},
  {Serenelli}, {McMillan}, {Lind}, {Bensby}, {Bergemann}, {Asplund},
  {Vallenari}, {Flaccomio}, {Pancino}, {Korn}, {Recio-Blanco}, {Bayo},
  {Carraro}, {Costado}, {Damiani}, {Heiter}, {Hourihane}, {Jofr{\'e}},
  {Kordopatis}, {Lardo}, {de Laverny}, {Monaco}, {Morbidelli}, {Sbordone},
  {Worley}, \& {Zaggia}}]{ruchti+15}
{Ruchti} G.~R. {et~al.}, 2015, \mnras, 450, 2874

\bibitem[{Rzhetsky \& Nei(1993)}]{Rzhetsky01091993}
Rzhetsky A., Nei M., 1993, Molecular Biology and Evolution, 10, 1073

\bibitem[{Saitou \& Nei(1987)}]{Saitou01071987}
Saitou N., Nei M., 1987, Molecular Biology and Evolution, 4, 406

\bibitem[{{Sanders} \& {Binney}(2016)}]{sanders+16}
{Sanders} J.~L., {Binney} J., 2016, \mnras, 457, 2107

\bibitem[{{Sch{\"o}nrich}(2012)}]{schonrich+12}
{Sch{\"o}nrich} R., 2012, \mnras, 427, 274

\bibitem[{{Sch{\"o}nrich} \& {Binney}(2009)}]{schonrich+09}
{Sch{\"o}nrich} R., {Binney} J., 2009, \mnras, 399, 1145

\bibitem[{{Sellwood} \& {Binney}(2002)}]{sellwood+02}
{Sellwood} J.~A., {Binney} J.~J., 2002, \mnras, 336, 785

\bibitem[{{Shen} {et~al}\mbox{.}(2016){Shen}, {Kulkarni}, {Madau}, \&
  {Mayer}}]{2016arXiv161202832S}
{Shen} S., {Kulkarni} G., {Madau} P., {Mayer} L., 2016, ArXiv e-prints

\bibitem[{Sneath(1973)}]{sneath1973numerical}
Sneath, P. H. A.and~Sokal R.~R., 1973, Numerical taxonomy. The principles and
  practice of numerical classification, ISBN: 0716706970. W.H. Freeman, San
  Francisco, US

\bibitem[{{Spina} {et~al}\mbox{.}(2016){Spina}, {Mel{\'e}ndez}, {Karakas},
  {Ram{\'{\i}}rez}, {Monroe}, {Asplund}, \& {Yong}}]{2016arXiv160604842S}
{Spina} L., {Mel{\'e}ndez} J., {Karakas} A.~I., {Ram{\'{\i}}rez} I., {Monroe}
  T.~R., {Asplund} M., {Yong} D., 2016, \aap, 593, A125

\bibitem[{Studier \& Keppler(1988)}]{Studier01111988}
Studier J.~A., Keppler K.~J., 1988, Molecular Biology and Evolution, 5, 729

\bibitem[{{Ting} {et~al}\mbox{.}(2012){Ting}, {Freeman}, {Kobayashi}, {De
  Silva}, \& {Bland-Hawthorn}}]{ting+12}
{Ting} Y.-S., {Freeman} K.~C., {Kobayashi} C., {De Silva} G.~M.,
  {Bland-Hawthorn} J., 2012, \mnras, 421, 1231

\bibitem[{{Tolstoy}, {Hill} \& {Tosi}(2009){Tolstoy}, {Hill}, \&
  {Tosi}}]{2009ARA&A..47..371T}
{Tolstoy} E., {Hill} V., {Tosi} M., 2009, \araa, 47, 371

\bibitem[{{van Leeuwen}(2007)}]{2007A&A...474..653V}
{van Leeuwen} F., 2007, \aap, 474, 653

\bibitem[{{Wielen}(1977)}]{wielen77}
{Wielen} R., 1977, \aap, 60, 263

\bibitem[{{Wojno} {et~al}\mbox{.}(2016){Wojno}, {Kordopatis}, {Steinmetz},
  {McMillan}, {Matijevi{\v c}}, {Binney}, {Wyse}, {Boeche}, {Just}, {Grebel},
  {Siebert}, {Bienaym{\'e}}, {Gibson}, {Zwitter}, {Bland-Hawthorn}, {Navarro},
  {Parker}, {Reid}, {Seabroke}, \& {Watson}}]{2016MNRAS.461.4246W}
{Wojno} J. {et~al.}, 2016, \mnras, 461, 4246

\bibitem[{Yang \& Rannala(2012)}]{yang2012molecular}
Yang Z., Rannala B., 2012, Nature Reviews Genetics, 13, 303

\end{thebibliography}
